\documentclass[sigconf]{acmart}
\usepackage{enumitem}
\usepackage{epstopdf}
\usepackage{multirow}
\usepackage{balance}
\usepackage[ruled,linesnumbered]{algorithm2e}
\usepackage{caption}
\usepackage{subcaption}
\usepackage{graphicx}
\usepackage{float}
\usepackage{adjustbox}
\usepackage{wrapfig,lipsum}
\usepackage{booktabs,etoolbox}
\newlength{\toprulewidth}
\patchcmd{\toprule}
  {\heavyrulewidth}{\toprulewidth}
  {}{}
\newcommand{\RN}[1]{%
  \textup{\uppercase\expandafter{\romannumeral#1}}%
}

\def\mbertmarco{m\textsuperscript{2}BERT}
\def\matbert{MART}
\def\matbertplb{MART-PLB}

\AtBeginDocument{%
  \providecommand\BibTeX{{%
    \normalfont B\kern-0.5em{\scshape i\kern-0.25em b}\kern-0.8em\TeX}}}

\copyrightyear{2021} 
\acmYear{2021} 
\setcopyright{acmcopyright}\acmConference[CIKM '21]{Proceedings of the 30th ACM International Conference on Information and Knowledge Management}{November 1--5, 2021}{Virtual Event, QLD, Australia}
\acmBooktitle{Proceedings of the 30th ACM International Conference on Information and Knowledge Management (CIKM '21), November 1--5, 2021, Virtual Event, QLD, Australia}
\acmPrice{15.00}
\acmDOI{10.1145/3459637.3482452}
\acmISBN{978-1-4503-8446-9/21/11}

\begin{document}
\fancyhead{}

\title{Mixed Attention Transformer for Leveraging Word-Level Knowledge to Neural Cross-Lingual Information Retrieval}

\author{Zhiqi Huang, Hamed Bonab, Sheikh Muhammad Sarwar, Razieh Rahimi, and James Allan}
\affiliation{
  \institution{Center for Intelligent Information Retrieval\protect\\University of Massachusetts Amherst}
  \country{}
}
\email{{zhiqihuang, bonab, smsarwar, rahimi, allan}@cs.umass.edu}


\begin{abstract}
Pre-trained contextualized representations offer great success for many downstream tasks, including document ranking. The multilingual versions of such pre-trained representations provide a possibility of jointly learning many languages with the same model. Although it is expected to gain big with such joint training, in the case of cross-lingual information retrieval (CLIR), the models under a multilingual setting are not achieving the same level of performance as those under a monolingual setting. We hypothesize that the performance drop is due to the \textit{translation gap} between query and documents. In the monolingual retrieval task, because of the same lexical inputs, it is easier for model to identify the query terms that occurred in documents. However, in the multilingual pre-trained models that the words in different languages are projected into the same hyperspace, the model tends to “translate” query terms into related terms – i.e., terms that appear in a similar context – in addition to or sometimes rather than synonyms in the target language. This property is creating difficulties for the model to connect terms that co-occur in both query and document.
To address this issue, we propose a novel Mixed Attention Transformer (MAT) that incorporates external word-level knowledge, such as a dictionary or translation table. We design a sandwich-like architecture to embed MAT into the recent transformer-based deep neural models. By encoding the translation knowledge into an attention matrix, the model with MAT is able to focus on the mutually translated words in the input sequence. Experimental results demonstrate the effectiveness of the external knowledge and the significant improvement of MAT-embedded neural reranking model on CLIR task.
\end{abstract}

\ccsdesc[500]{Information systems~Information retrieval}
\ccsdesc[500]{Information systems~Multilingual and cross-lingual retrieval}
\ccsdesc[300]{Information systems~Retrieval models and ranking}

\keywords{Cross-lingual information retrieval; Attention mechanism; Neural network}

\maketitle
\section{Introduction}  \label{sec:introduction}

We study the problem of Cross-Lingual Information Retrieval (CLIR) in which the desired information is written in a language different than that of the user’s query. From the modeling perspective, in the CLIR setting some form of language translation is needed to map  the vocabulary of the query language to that of the documents’ language in addition to the ranking component. 
This translation gap can be bridged with simple dictionaries, translation tables, machine translation, or more recently cross-language distributional representations~\cite{yarmohammadi-etal-2019-robust,bonab2020training, sarwar-etal-2019-multi}.

Embedding the translation component in the fine-tuning stage along with the ranking makes the training of deep neural models for the CLIR more challenging, particularly when dealing with resource-lean languages~\cite{10.1145/3341981.3344236,10.1145/3331184.3331324}. 
Pre-trained language  models such as BERT~\cite{devlin-etal-2019-bert} have shown promising performance gains for monolingual information retrieval~\cite{jiang-etal-2020-cross, qiao2019understanding,yates-etal-2021-pretrained,10.1145/3397271.3401325}. This success is mainly due to the unsupervised pre-training of context-aware transformer architectures with an enormous number of parameters over large corpora. To achieve success in the learning-to-rank task such models are often fine-tuned with a relatively large collection of relevance judgments such as the MS MARCO  passage ranking dataset~\cite{nogueira2019passage}.
However, it is not feasible to obtain data in the scale of MS MARCO across different languages. 
Thus, some studies leverage different training data (e.g., weak-supervised data, cross-lingual Wikipedia-based data~\cite{sasaki2018cross,schamoni2014learning}) and techniques (e.g., domain adaptation, few-shot learning) in order to adapt the model for the target task and language, reporting improvements. 

The multilingual versions of pre-trained Transformer-based language models, such as mBERT \cite{devlin-etal-2019-bert} and XLM-R \cite{conneau2020unsupervised}, provide the possibility of jointly learning representations for multiple languages with the same model. Fine-tuning these pre-trained multilingual language models for ranking, similar to the monolingual setting, enables cross-language information retrieval. 
In the multilingual pre-trained models that words in different languages are projected into the same hyperspace, the model tends to map query terms into target language's related terms – i.e., terms that appear in a similar context – in addition to or sometimes rather than synonyms~\cite{qiao2019understanding, bonab2020training}. We hypothesize that this phenomena creates difficulties for the model to connect terms that match between the query and document. It has been shown that the translation gap plays a significant role in the suboptimal success of neural CLIR models and addressing that can significantly boost the retrieval performance~\cite{bonab2020training}. 
Therefore, the multilingual language models for the CLIR task have not yet achieved the performance gain observed with the use of pre-trained language models for monolingual information retrieval~\cite{10.1145/3442381.3449830, litschko2018unsupervised}. 
This can happen in the CLIR task because the vocabulary size is almost doubled, the possibility of exact match  between query and document is limited, and training data (e.g., bilingual query log or click data) is scarce.
Most of the existing CLIR systems are thus deployed along with a query translation component to reduce the problem into monolingual retrieval. However, it is important to note that having a translation component as a black-box limits the retrieval component due to translation errors. 



We inject word-level translation knowledge into a model at the time of fine-tuning it with relevance data.  More specifically, we leverage the external knowledge in the form of a translation table, which is a look-up table that provides translation probabilities for a pair of words in two different languages. We use the translation table to create an attention matrix and use it in parallel with the Transformer’s multi-head attention -- both in our training and inference phase -- to improve the model's cross-lingual understanding. We refer to our extended component as Mixed Attention Transformer (MAT) and create MART, a sandwich-like architecture to embed MAT into the multilingual BERT (mBERT) model. By encoding the translation knowledge into an attention matrix, we enable the overall architecture to focus on the mutually translated words in the input sequence. Our experiments explore the effectiveness of a variety of external knowledge sources and show the significant gain that we get from MART on CLIR task. MAT is a generalized architecture capable to capture any form of lexical mapping and it can be integrated with any transformer-based architecture.



We performed extensive experiments on ten different language pairs for CLIR training and evaluation, three different resources to obtain the translation knowledge, and different qualities of translations based on available translation resources for language pairs.
Our experimental results demonstrate the varied effectiveness of different external knowledge sources and the significant improvement of MAT-embedded neural re-ranking model over strong baselines on the CLIR task.
In terms of mean average precision (MAP), our proposed model outperforms the neural baseline by 8\% on high-resource languages and 12\% on low-resource languages.

The rest of this paper is organized as follows. In Section 2 we provide a review of related works. Section 3 presents our MAT architecture for injecting external translation knowledge directly into model. Section 4 and 5 provides our experimental design and results with discussions and further analyses. We conclude our study in Section 6.

\section{Related Work}  \label{sec:related-work}

We first provide a summary of existing CLIR models trained from both word-embedding based representations as well as representations from unsupervised language models based on the transformer architecture. We discuss the importance of the knowledge from sentence-level parallel data and how they enhance the performance of neural retrieval models. Finally, we also elaborate on the transformer-based architectures that incorporate external knowledge for a variety of tasks and compare them to MAT.

\subsection{Neural Cross-lingual Representation Spaces}

CLIR tackles two sub-tasks: query translation and query-document matching, and neural models are applicable to both the tasks. One approach is to translate the query to the language of the corpus by using a Statistical Machine Translation (SMT) or Neural Machine Translation (NMT) model and then apply a mono-lingual matching model to determine the relevance. While the translate-then-retrieve approach is a popular one, neural bilingual word representations creates the opportunity to skip the translation step. As a result, query-document matching can  performed in a shared vector space for two languages, where similar words in two different languages are mapped close to each other. The assumption is Cross-linual Word Embeddings (CLWE) are capable to bridge the translation gap between two languages. 

One of the earliest works in this direction is from \citet{vulic2015monolingual}, and they proposed a model to learn bilingual word embeddings using a document-aligned comparable data. Once all the words in both languages are represented in a shared space, they computed query and document representations using the compositional distributional semantics model and calculated their matching score based on cosine similarity metric. \citet{litschko2018unsupervised} used the same matching technique but created the shared space using only monolingual data in two languages. \citet{bonab2020training} assessed the effectiveness of several bilingual word embeddings under cosine similarity-based scoring framework for retrieval and found that all the existing word embeddings lack the capacity to translate a source language word into the target language word -- they refer to this phenomenon as the \emph{translation gap}. The authors showed that a bilingual word embedding brings similar pair of words in two languages close together, but often keeps the words that are translation of each other far than expected. This is because cross-lingual word embeddings are learned from the contextual information around a word but not from the translation of that word. The authors proposed a smart shuffling approach to include translation knowledge into word embeddings and created a state-of-the-art cross-lingual word embedding for retrieval. While it is clear that translation knowledge brings significance gain in retrieval, there is no study on how to incorporate this knowledge in the modern transformer based query-document matching frameworks. 

Unsupervised multilingual language models based on the transformer architecture (also referred to as multilingual transformers) brought a major advancement over the cross-lingual word embeddings. There are two major realization of such models: mBERT (Multilingual BERT)~\cite{devlin-etal-2019-bert} and XLM-R (XLM RoBERTa)~\cite{conneau-etal-2020-unsupervised}. These models offer a shared representation space for a large number of languages and the representation of a token is contextualized based on the other tokens in a sequence. Thus these approaches capture higher-level semantics compared to CLWE and once fine-tuned, they have been shown to be effective across a wide variety of tasks, including CLIR~\cite{saleh-pecina-2020-document,10.1145/3331184.3331324, 10.1145/3397271.3401322}. However, we assume that the \emph{translation gap} still exists in the multilingual transformers and it is important to inject translation knowledge into such architectures. 

\subsection{Neural Matching Models for CLIR}
Whether we use cross-lingual word embeddings or multilingual transformers for representing query and documents, we need to provide relevance knowledge to these models for effective matching. Thus, we need to further train these language representation spaces using with relevance judgments from human~\cite{sasaki2018cross, zhao2019weakly, li2018learning, bonab2020training}. 

\citet{sasaki2018cross} constructed a large-scale weakly supervised CLIR collection by using the first sentence of a Wikipedia page as the query and all the linked foreign language articles as documents. They proposed a shallow learing-to-rank method and did not use a shared language representation space. Thus, their approach does not explicitly close the language gap between the query and document. \citet{zhao2019weakly} leverages the sentence-aligned parallel data to create weakly supervised relevance judgments. They use a sentence from a language as a document and randomly select a word from the translation of that sentence as query. Even though they close the language gap using parallel data, they do not use relevance judgments explicitly. We use both parallel data and relevance judgments and improve the architecture of a multilingual transformer to adapt these sources. 

Rather than considering relevance and translation in isolation, \citet{li2018learning} took an adversarial learning approach to jointly learn language alignment through translation knowledge and cross-lingual matching using relevance judgments. They created a weakly-supervised collection of parallel data by translating AOL queries using Google Translate. They use a Long Short Term Memory (LSTM) network to learn matching in contrast to the multilingual transformer proposed in this work. Moreover, they use weak parallel data to close the language gap, whereas we use word-level alignment learned from the parallel data or obtained from a dictionary in the fine-tuning stage. \citet{bonab2020training} achieved state-of-the-art performance when they used their translation-oriented bilingual representations with DRMM matching model~\cite{Guo_2016} and trained the architecture using relevance data. They showed that dictionary-oriented word embeddings can improve the performance of a DRMM model when fine-tuned with relevance data. We propose a novel multilingual transformer architecture, MAT, which learns jointly from relevance judgments and translation knowledge in the form of a dictionary or a translation table.

\subsection{Knowledge Injection into Transformers}
There has been a number of efforts to inject structured world
knowledge into unsupervised pretraining and contextualized representations \cite{zhang-etal-2019-ernie, Lauscher2020SpecializingUP, Levine2020SenseBERTDS, Peters2019KnowledgeEC, He2020IntegratingGC, Xiong2020PretrainedEW}. Most of these works focus on integrating knowledge-graphs information such as type of an entity or relatedness between a pair of entities. \citet{Lauscher2020SpecializingUP} incorporated lexical semantics into BERT by injecting word pairs that are synonyms or hold hyponym-hypernym relations in WordNet. \citet{Levine2020SenseBERTDS} injected word-supersense knowledge by predicting the supersense of a masked word in the input and the ground truth is obtained from Wikipedia. All these works augment an extra knowledge-driven loss with the standard language modeling loss in the language model pre-training stage. We augment translation knowledge in the form of attention in the fine-tuning stage. Our approach is flexible as we can adapt new knowledge as more data for fine-tuning becomes available. 

A recent work from \citet{attention_sts} used attention-based approach to integrate lexical knowledge for the semantic textual matching task. They created an attention matrix from WordNet and computed the Hadamard product of the attention matrix with BERT's attention matrix. They investigated this approach for computing sentence similarity in a monolingual setting.
\section{Methodology}  \label{sec:methodology}



Our goal is to incorporate additional knowledge from statistical machine translation models or human-constructed dictionaries into a transformer architecture to enable it to connect query and document tokens -- not only based on relevance, but also based on translations.
In this section, we first define the translation attention matrix given an input query and a candidate document. Then we introduce the translation attention head and the Mixed Attention Transformer (MAT) layer. Finally, we design a sandwich-like architecture to embed MAT into the existing transformer-based neural ranking model.

\subsection{Translation Attention Matrix}
We define translation reference as a large structural dataset containing knowledge to translate words from one language to another e.g, a human-constructed dictionary, or a translation table trained on parallel corpora. 
In the CLIR task, the  translation knowledge is dependent on the query and document.
Therefore, we first design an algorithm that distill the translation knowledge based on tokens in the query and document.

Suppose there exists a word-level translation reference $T$. Given word $w_{s}$ in the source language and $w_{t}$ in the target language, $T(w_t, w_s)$ returns the probability of $w_s$ being translated to $w_{t}$: $T(w_t, w_s) = P(w_t|w_s, T)$.


We assume the query is in the source language with length of $m_q$ words and the document is in the target language with length of $m_d$ words. Therefore, the concatenation of query and document $[q, d]$ has length $m = m_q + m_d$.
Then we construct a $m \times m$ translation attention matrix $M^{tr}$ based on $[q, d]$ and $T(\cdot,\cdot)$ by symmetrically assigning translation probabilities between query tokens and document tokens. We provide detailed instructions for constructing $M^{tr}$ in Algorithm \ref{alg:trans-mat}.

\begin{algorithm}[t]
\SetKwInput{KwInput}{Input}
\SetKwInput{KwOutput}{Output}
\SetKwInput{KwReturn}{return}
\DontPrintSemicolon
  \KwInput{$[q, d]$ and $T(\cdot,\cdot)$}
  \KwOutput{$M^{tr}$}
  Initialize $M^{tr}$ as a $m \times m$ zero matrix.\;
  \For{each token $w_k$ in the input sequence}    
  { 
    	$M^{tr}_{kk} = 1$
  }
  \For{each query token $w_i$}    
  { 
    \For{each document token $w_j$}
    {
        $M^{tr}_{ij} = M^{tr}_{ji} = T(w_j, w_i)$
    }
  }
  $M^{tr} \leftarrow \mathrm{RowNorm}(M^{tr})$\;
  \KwReturn{$M^{tr}$}
\caption{Generate translation attention matrix}
\label{alg:trans-mat}
\end{algorithm}

Note that the $k$\textsuperscript{th} row of $M^{tr}$ represents the attention weights of $k$\textsuperscript{th} token in the input assigned across all the input tokens. In Algorithm~\ref{alg:trans-mat}, lines 2-4 guarantee each token, including out-of-vocabulary word, is assigned a weight to itself and the self weight is the upper bound of all of its translation probabilities. If $q_i$ and $d_j$ are mutually translated words, they get their translation probabilities to each other from lines 5-9. Finally, the row normalization ensures that the attention weights for each input token sum up to 1.

\begin{figure}[t]
    \captionsetup[subfigure]{font=footnotesize,labelfont=footnotesize}
    \begin{subfigure}[t]{0.5\textwidth}
        \centering
        \includegraphics[width=0.9\linewidth]{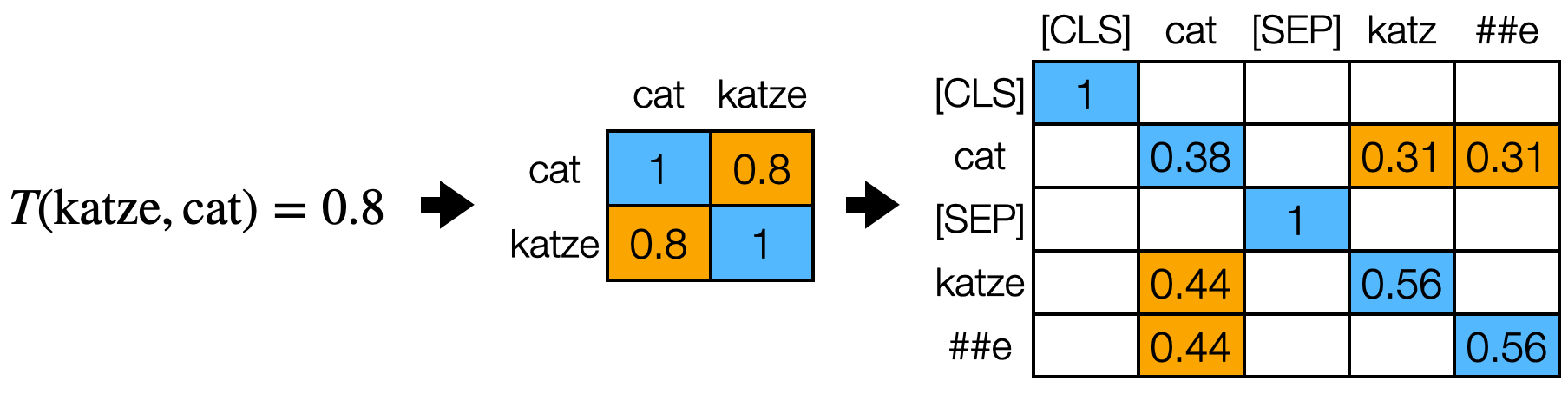}
    \end{subfigure}
\caption{A toy example for generating $M^{tr}$.}
\label{fig:mtr-example}
\end{figure}

To encode rare words with limited vocabulary size, Byte Pair Encoding (BPE) is often used by pre-trained language models, which splits words into sub-word units. 
There is evidence that self-attention treats split words differently than non-split ones~\cite{correia-etal-2019-adaptively}. Therefore, we use tokens before BPE to query the translation reference and then assign the same attention weight to all parts of the same word. The dimension $m$ of $M^{tr}$ is the same as the length of sequence of $[q, d]$ tokenized by a pre-trained language model. A simplified example for generating $M^{tr}$ with query ``cat'' and document ``katze'' (German translation of cat) is shown in Figure~\ref{fig:mtr-example}.


\subsection{Mixed Attention Transformer}
In order to inject $M^{tr}$ into  a transformer-based model, we propose a novel transformer network, named Mixed Attention Transformer (MAT) by combining the multi-head attention with translation-based attention. 

The multi-head attention~\cite{vaswani2017attention} is the core of the transformer architecture which consists of $n$ different attention heads. Given the vector representations as the hidden states $\mathbf{h}$, each head computes the dot-product attention:
$$
\mathrm{Attention}_{i}(\mathbf{h}) = \mathrm{softmax}\Big(\frac{W_i^{q}\mathbf{h}\cdot W_i^{k}\mathbf{h}}{\sqrt{d/n}}\Big)W_i^{\nu}\mathbf{h}
$$
where $\mathbf{h}$ is a $d$ dimensional hidden vector for an input sequence. In BERT, the $W_i^{q}$, $W_i^{k}$ and $W_i^{\nu}$ are matrices with size $d/n \times d$. Thus, each head projects to a different subspace of size $d/n$, learning different information.


Then the outputs of the multi-head attention, $\mathrm{MH}(\cdot)$, are concatenated $n$ heads together and linearly transformed:
$$
\mathrm{MH}(\mathbf{h}) = W^{o}[\mathrm{Attention}_{1}, \ldots, \mathrm{Attention}_{n}]
$$
In parallel to multi-head attention, we  introduce the translation attention head denoted as $\mathrm{TH}(\cdot)$. Inspired by the scaled dot-product attention, we replace the attention weights learned from matrices $W_i^{q}$ and $W_i^{k}$ by the fixed attention weights in $M^{tr}$. Then, the multi-head attention becomes a single fixed attention head as follows
\begin{equation*}
    \mathrm{TH}(\mathbf{h}) = W_{\mathrm{TH}}^o\big(M^{tr}(W_{\mathrm{TH}}^{\nu}\mathbf{h})\big),
\end{equation*}
where both $W_{\mathrm{TH}}^o$ and $W_{\mathrm{TH}}^\nu$ are trainable matrices in $\mathrm{TH}(\cdot)$ with dimension $d \times d$.
By matrix multiplying with $M^{tr}$, the translation attention head is capable to reduce the distance between mutually translated tokens in the token representation hyperspace. We prove the effect of $M^{tr}$ on hidden states in a simplified scenario. 

\noindent\textbf{Lemma 1.} Let convex combinations of vectors A and B be $\alpha A + \beta B$ and $\beta A + \alpha B$ where $\alpha + \beta = 1$. Then, the cosine similarity between  $\alpha A + \beta B$ and $\beta A + \alpha B$ is greater or equal to the cosine similarity between $A$ and $B$.

\noindent\textbf{Proof.} 
\begin{align*}
    Sim(\alpha A + \beta B, \beta A + \alpha B) & = \frac{(\alpha A + \beta B)\cdot (\beta A + \alpha B)}{\lVert \alpha A + \beta B \rVert \lVert \beta A + \alpha B \rVert}\\
    & \geq \frac{(\alpha^2 +\beta^2)A\cdot B +\alpha\beta(\lVert A \rVert^2 +\lVert B \rVert^2)}{(\alpha^2 +\beta^2)\lVert A \rVert\lVert B \rVert + \alpha\beta(\lVert A \rVert^2 +\lVert B \rVert^2)} \\
    & \geq \frac{A\cdot B}{\lVert A \rVert\lVert B \rVert}.
\end{align*}
\noindent Therefore, $Sim(\alpha A + \beta B, \beta A + \alpha B) \geq Sim(A, B)$.

Suppose query word $w_i$ and document word $w_j$ are the translations of each other with probability $p > 0$, and words other than $w_j$ in documents all have zero translation probability with $w_i$. Then, the only two non-zero weights in the $i$\textsuperscript{th} row of $M^{tr}$ are self attention ($M_{ii}^{tr}$) and attention on $w_j$ ($M_{ij}^{tr}$):
$$
M_{ii}^{tr} = \frac{1}{(1+p)}; \ M_{ij}^{tr} = \frac{p}{(1+p)}
$$
Similarly for $w_j$, the non-zero weights in the $j$\textsuperscript{th} row are $M_{jj}^{tr} = 1/(1+p)$ and $M_{ji}^{tr} = p/(1+p)$. If we ignore the trainable matrices in $\mathrm{TH}(\cdot)$ and directly multiply $M^{tr}$ with hidden states $\mathbf{h}$, the translation attention output of $w_i$ and $w_j$ are a convex combination of each other's hidden representations:
$$
\mathrm{TH}(\mathbf{h}_{w_i}) = \frac{1}{1+p}\mathbf{h}_{w_i} + \frac{p}{1+p}\mathbf{h}_{w_j}
$$
$$
\mathrm{TH}(\mathbf{h}_{w_j}) = \frac{1}{1+p}\mathbf{h}_{w_j} + \frac{p}{1+p}\mathbf{h}_{w_i}
$$
According to \textbf{Lemma 1}, because $p > 0$,
$$
Sim\big(\mathrm{TH}(\mathbf{h}_{w_i}), \mathrm{TH}(\mathbf{h}_{w_j})\big) > Sim(\mathbf{h}_{w_i}, \mathbf{h}_{w_j})
$$
Thus, when $p$ is large, the words in query and document are likely to be translation to each other. The attention matrix $M^{tr}$ ``pays attention'' to all these pair of words and $\mathrm{TH}(\cdot)$ tends to ``pull'' their hidden representations closer in the hyperspace.

\begin{figure}[t]
    \captionsetup[subfigure]{font=footnotesize,labelfont=footnotesize}
    \begin{subfigure}[t]{0.5\textwidth}
        \centering
        \includegraphics[width=0.98\linewidth]{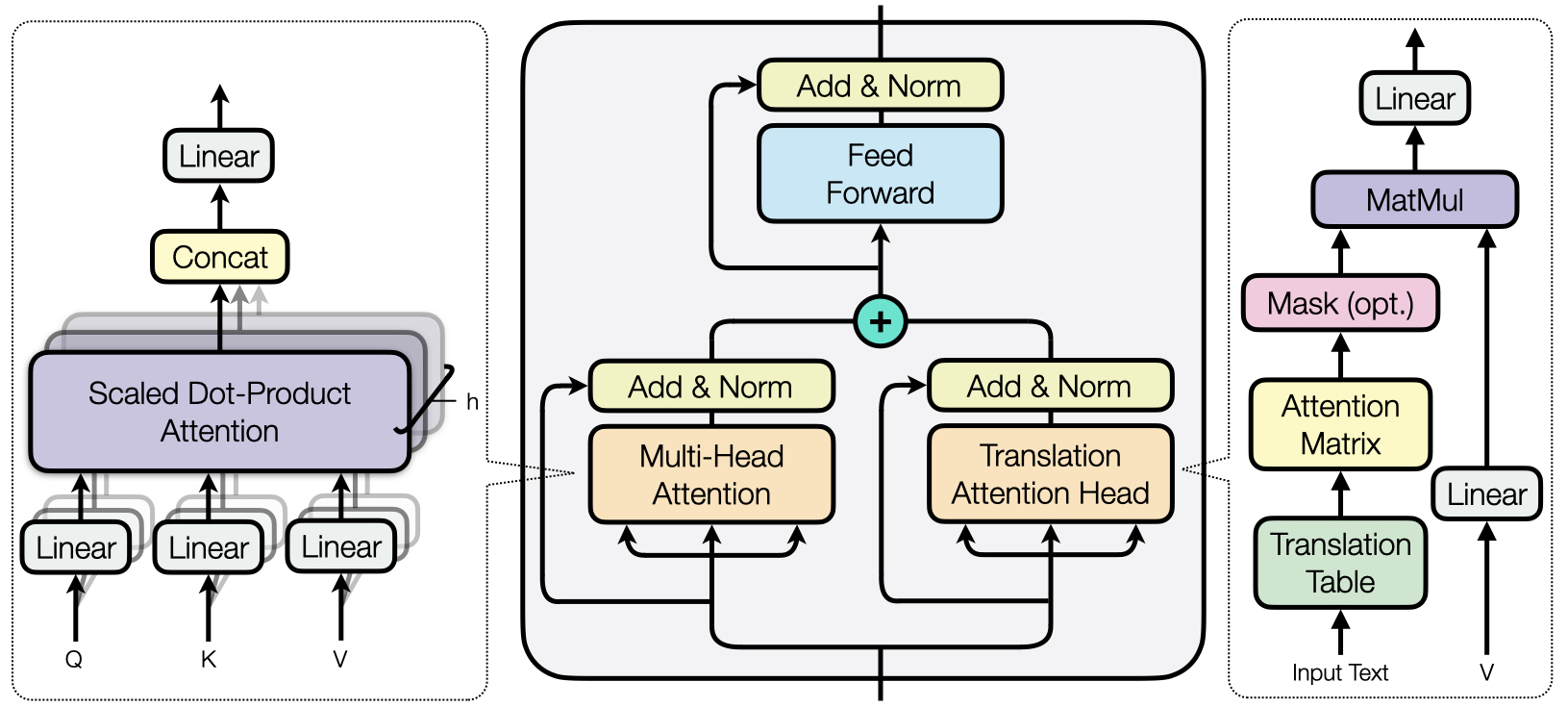}
    \end{subfigure}
\vspace{-0.3cm}
\caption{(left) Multi-Head Attention. (right) Translation Attention Head. (middle) Mixed Attention Transformer Layer.}
\vspace{-0.1cm}
\label{fig:model}
\end{figure}

The complete attention mechanism in MAT is a combination of the attention outputs from both $\mathrm{MH}(\cdot)$ and $\mathrm{TH}(\cdot)$.
We first employ a residual connection around each type of attention output, followed by layer normalization, denoted as $\mathrm{LN}(\cdot)$, resulting two sub-layer outputs. Then we sum two sub-layer outputs:
\begin{gather*}
    \mathrm{Sublayer}_{\mathrm{MH}}(\mathbf{h}) = \mathrm{LN}(\mathbf{h} + \mathrm{MH}(\mathbf{h})) \\
    \mathrm{Sublayer}_{\mathrm{TH}}(\mathbf{h}) = \mathrm{LN}(\mathbf{h} + \mathrm{TH}(\mathbf{h}))\\
    \mathbf{h}' = \mathrm{Sublayer}_{\mathrm{MH}}(\mathbf{h}) + \mathrm{Sublayer}_{\mathrm{TH}}(\mathbf{h})
\end{gather*}
And apply the summed result to the position-wise feed-forward networks (FFN), 
\begin{gather*}
    \mathrm{FFN}(x) = \max(0, xW_1 + b_1 )W_2 + b_2
\end{gather*}
The final output of MAT is another residual connection around the output of FFN:
\begin{gather*}
    \mathrm{MAT}(\mathbf{h}) = \mathrm{LN}(\mathbf{h}' + \mathrm{FFN}(\mathbf{h}'))
\end{gather*}
The complete MAT architecture is depicted in Figure~\ref{fig:model} (middle). The left and right of Figure~\ref{fig:model} are two types of attention component in MAT. The benefits of this network architecture are that the MAT can attend to both contextual information from multi-head attention and translation knowledge from translation attention head during training. Because we keep the multi-head attention mechanism and share the FFN sublayer, MAT contains a vanilla transformer network. This design allows MAT to be easily embedded into recent transformer-based pre-trained models and fully leverage the pre-trained weights.

\subsection{Embed MAT into Pre-trained Model}

The transformer-based models usually have the following architecture: First, the embedding layer encodes the input tokens, segments, and positions into hidden representations. 
The representation of each input token is then updated by a stack of encoder layers based on the attention mechanism.
Finally, a specialized add-on network maps the hidden representations to an output based on the task.

\citet{qiao2019understanding} analyzed different ranking models based on BERT and found that the \texttt{Last-Int} approach which applies BERT on the concatenated $[q,d]$ sequence and uses the last layer’s representation of the [CLS] token as the matching feature gives the best performance. In this section, we use the same BERT (\texttt{Last-Int}) as a re-ranker to discuss how to embed MAT into a transformer-based pre-trained language model. 

\matbert{} (\textbf{MA}T+BE\textbf{RT}), the new model architecture we propose is to keep  the embedding layer and add-on network while replacing some of the transformer layers in the middle by MAT. 

\begin{figure}[t]
    \captionsetup[subfigure]{font=footnotesize,labelfont=footnotesize}
    \begin{subfigure}[t]{0.3\textwidth}
        \centering
        \includegraphics[width=0.75\linewidth]{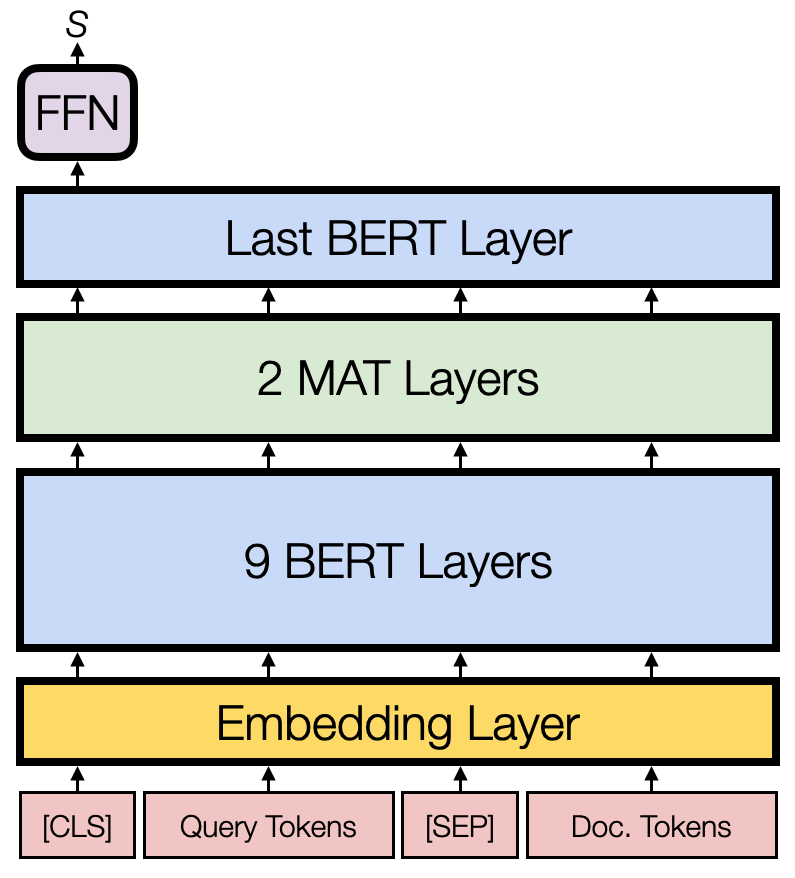}
    \end{subfigure}
\vspace{-0.3cm}
\caption{Use MAT layers in BERT ranking model}
\vspace{-0.1cm}
\label{fig:ranker}
\end{figure}

During fine-tuning, the BERT layers close to the output (higher layers) are more sensitive than the lower layers~\cite{zhao-bethard-2020-berts}. 
Also, another study on BERT~\cite{tenney-etal-2019-bert} has shown  that most local syntactic phenomena are encoded in lower layers while higher layers capture more complex semantics. Consider the fine-tuning efficiency and semantic quality of the token representations, the layer replacement should start from the higher layers of BERT. 
Moreover, in the \texttt{Last-Int} ranking approach, the output score is only based on the [CLS] token in the last BERT layer.
Therefore, we keep the last BERT (\texttt{Base}) layer as the output layer and start to embed MAT from the 11\textsuperscript{th} layer. Figure~\ref{fig:ranker} shows an examples of the sandwich-like architecture based on a BERT-based ranking model where MAT layers are embedded into 10\textsuperscript{th} and 11\textsuperscript{th} layers of BERT. 
Using the same hidden dimension as BERT, each MAT layer only introduces about 1.18M new parameters comparing to the BERT layer. At initialization, MAT is able to use pre-trained weights of its corresponding BERT layer. This compatibility increases the fine-tuning efficiency and reduces the training data requirement.


\section{Experimental Setup}  \label{sec:experiments}
\subsection{Dataset}
\textbf{CLIR dataset.} We create our training and evaluation data from the Cross-Language Evaluation Forum (CLEF) 2000-2008 campaign for bilingual ad-hoc retrieval tracks~\cite{10.1007/3-540-44645-1_9, 10.1007/3-540-45691-0_2, braschler2002clef, braschler2003clef, 10.1007/11519645_1, 10.1007/11878773_1, 10.1007/978-3-540-74999-8_1, 10.1007/978-3-540-85760-0_1, 10.1007/978-3-642-04447-2_1}. We use the text fields of the documents to construct our retrieval corpus and discard other meta data. We concatenate the title and description fields of a topic and consider it as our query. 
We consider all the topics and relevance judgments from all the tracks to show the consistent effectiveness of MAT across several cross-language retrieval settings on both high- and low-resource languages.

\textbf{Translation Resources.} 
Our goal is to leverage translation resources as external knowledge into the query-document matching process and we compare the effectiveness of three types of resources: sentence-level parallel data, dictionary, and bi-lingual word embeddings. We use sentence-level parallel data with GIZA++ toolkit~\cite{och2003systematic} to construct a translation table, which we use to generate $M^{tr}$. Translation tables for European languages are based on the Europarl v7 sentence-aligned corpora~\cite{koehn2005europarl}. For our limited-resource (both in terms of parallel data and relevance judgments) setting based on Somali and Swahili languages, we use the translation tables provided by~\citet{zhang-etal-2020-2019}.


As the dictionary-based translation resource we use Panlex, a dictionary~\cite{kamholz2014panlex} whose data acquisition strategy emphasizes high-quality lexical mapping and broad language coverage. Finally, we also explore the a multilingual word embedding as a translation resource following \citet{bonab2020training}. Given a pair of words we use their representations from a multilingual word embedding model and compute cosine similarity to model relatedness of the pair of words. In our experiments, we use MUSE, an unsupervised multilingual word embedding from~\cite{conneau2017word} as translation resource.  

\textbf{Text Pre-processing.} In order to have consistent pieces of text across different resources, we normalize characters by mapping diacritic characters to the corresponding unmarked characters and then lower-casing text. For initial step of retrieval and translation table extraction from parallel corpora, we remove non-alphabetic, non-printable, and punctuation characters. We use NLTK library to tokenize and remove stop-words, but do not stem the tokens. 

\subsection{CLIR Settings}
\textbf{Forward: Non-English Query and English Documents.} In this setting, we use non-English queries against an English document collection. To evaluate cross-lingual matching performance, we use human translation of a fixed query set to obtain queries in different languages. While we have translations of queries in different languages, we keep the content and language of the retrieval corpus fixed. We have both high-resource and low-resource CLIR settings in our experiments. In a high-resource setting, for example, French-English, we have higher amount of sentence-level parallel data and relevance judgments compared to a low-resource setting.   



There are four high-resource language pairs in our experiments: French~(Fre-Eng), Italian~(Ita-Eng), German~(Deu-Eng), and Spanish~(Spa-Eng). Queries are selected from CLEF C001 -- C350 topic set for each language. We take the  intersection of the topic ID and remove topics without any relevant document, resulting in 246 overlapped queries across four languages. 
For cross-language information retrieval involving low-resource languages, we experiment on Somali~(Som-Eng) and Swahili~(Swa-Eng). ~\citet{10.1145/3341981.3344236} provided Somali and Swahili translations of 151 English queries from the CLEF C001 -- C200 topic set and we use those queries in our setting. The collection of English documents is the Los Angeles Times corpus comprised of 113k news articles.

\textbf{Backward: English Query and Non-English Documents} In this setting, we use English queries against document collections in four languages: French~(Eng-Fre), Italian~(Eng-Ita), German~(Eng-Deu) and Spanish~(Eng-Spa). For each language, we create a retrieval corpus from a combination of sources which we report in Table~\ref{tab:settings}.As the retrieval corpus varies for each language, relevance judgments are not available for all the English topics from CLEF C001 -- C350 topic set. Thus, for each CLIR setting we have a different number of queries in the backward setting compared to the forward setting. Table~\ref{tab:settings} provides information about query sets and document collections in both the settings.

\begin{table}[t]
    \centering
    \captionsetup{width=\linewidth}
    \caption{Summary of CLIR setting. First four rows indicate the backward and the last row indicates the forward setting.}
    \vspace{-0.3cm}
    \label{tab:settings}
    \begin{adjustbox}{width=0.48\textwidth}
    \begin{tabular}{lccc}
        \toprule
        CLIR Setting & Collection Source & Collection Size & Query Size\\
        \midrule
        \\[-1em]
        Eng-Fre & Le Monde, Sda French & 129,689 & 185\\
        Eng-Ita & La Stampa, Sda Italian & 144,040 & 176\\
        Eng-Deu & Der Spiegel, Frankfurter Rundschau & 153,496 & 184\\
        Eng-Spa & EFE News 94-95 & 452,027 & 156\\
        \midrule
        Xxx-Eng & Los Angeles Times 94  & 113,005 & 246\\
        \bottomrule
    \end{tabular}
    \end{adjustbox}
\end{table}

\subsection{Implementation Details}
\label{sec:implementation_details}
\textbf{Pre-trained passage re-ranker} \citet{nogueira2019passage} fine-tuned the \texttt{Base}, \texttt{Uncased} multilingual BERT (mBERT) on MS MARCO document retrieval dataset to create a passage ranking model. We refer to this pre-trained model as \mbertmarco{} and further fine-tune it with cross-lingual relevance judgments. 
To prepare the input sequence for \mbertmarco{} we concatenate a query and a document separated by a special [SEP] token from mBERT's vocabulary. We prefix the concatenated sequence with the special [CLS] token from mBERT's vocabulary. We obtain the last layer representation of this sequence from \mbertmarco{}, but only use the representation of the [CLS] token, and pass it through a linear combination layer to obtain the probability of the document being relevant to the query. At test time, given a query, \mbertmarco{} computes the probability for each document independently and obtains a document ranking after sorting with these probability scores. Because the mBERT input sequence is limited to 512 tokens, longer documents are split evenly and [CLS] representations from all document segments are averaged to obtain a representation for fine-tuning.  \citet{macavaney2019cedr} used the same approach for monolingual retrieval. 

\textbf{Evaluation.} For evaluating retrieval effectiveness, we follow prior work on CLEF dataset~\cite{10.1145/3331184.3331324, bonab2020training} and report mean average precision (MAP) of the top 100 ranked documents and precision of the top 10 retrieved documents (P@10). We determine statistical significance using the two-tailed paired \textit{t}-test with p-value less than 0.05 (i.e., 95\% confidence level).

\textbf{Model training.}
We train all neural re-ranking models using pairwise cross-entropy loss~\cite{dehghani2017neural}. We use all the positive document from the query relevance judgments and randomly sample negative documents to form training pairs. 
We truncate document contents to the first 800 tokens and create two passages to represent a document if the sum of the query length and document length is over the 512 tokens, which is the limit of mBERT. We pass a two query-document pairs in each forward pass but use gradient accumulation to make our effective batch size to 16. We train all the models for 100 epochs with an early stopping strategy with patience value of 20. All models are trained using Adam's optimization algorithm~\cite{kingma2014adam} with a learning rate of 2e-5.

Given the limited number of queries in each language, we use 5-fold cross-validation for robust evaluation. For each fold, the training, validation, and test data are 60\%, 20\%, and 20\% of the query set, respectively. The reported evaluation metrics are averaged across 5 folds. We also fix the random seed is set to guarantee that all models receive the same training data. For the validation queries, we re-rank the top 100 documents and use MAP to select the best-performing model. 

\begin{table*}[t]
    \centering
    \captionsetup{width=\linewidth}
    \caption{Model performance on forward and backward settings for high-resource languages. The highest value for each column is marked with bold text. Statistically significant improvements are marked by $\dag$ (over SMT) and $\ddag$ (over BERT).}
    \label{tab:main-comparison}
    \begin{adjustbox}{width=0.8\textwidth}
    \aboverulesep=0ex
    \belowrulesep=0ex
    \renewcommand{\arraystretch}{1.2}
    \begin{tabular}{c|lcccccccc}
        \toprule
        \multirow{8}{*}{\shortstack[c]{\textbf{Forward}\\\textbf{Setting}}} & \multirow{2}{*}{\textbf{Model}} & \multicolumn{2}{c}{Fre-Eng} & \multicolumn{2}{c}{Ita-Eng} & \multicolumn{2}{c}{Deu-Eng} & \multicolumn{2}{c}{Spa-Eng} \\
        \cmidrule(lr){3-4} \cmidrule(lr){5-6} \cmidrule(lr){7-8} \cmidrule(lr){9-10} 
        & & MAP & P@10 & MAP & P@10 & MAP & P@10 & MAP & P@10 \\
        \cmidrule{2-10}
        & Human Translation & $0.4569$ & $0.3940$ & $0.4569$ & $0.3940$ & $0.4569$ & $0.3940$ & $0.4569$ & $0.3940$ \\
        \cmidrule{2-10}
        & SMT & $0.3618$ & $0.3492$ & $0.3561$ & $0.3431$ & $0.3588$ & $0.3354$ & $0.3624$ & $0.3317$ \\
        \cmidrule{2-10}
        & \mbertmarco{} & $0.3802^{\dag}$ & $0.3799^{\dag}$ & $0.3652$ & $0.3545$ & $0.3582$ & $0.3335$ & $0.3819^{\dag}$ & $0.3693^{\dag}$ \\
        & \matbertplb{} & $0.3859^{\dag}$ & $0.3666^{\dag}$ & $0.3701$ & $0.3689^{\dag}$ & $0.3593$ & $0.3501^{\dag}$ & $0.3824^{\dag}$ & $0.3676^{\dag}$ \\
        & \matbert{} & $\mathbf{0.4126}^{\dag\ddag}$ & $\mathbf{0.3935}^{\dag\ddag}$ & $\mathbf{0.3944}^{\dag\ddag}$ & $\mathbf{0.3732}^{\dag\ddag}$ & $\mathbf{0.3862}^{\dag\ddag}$ & $\mathbf{0.3770}^{\dag\ddag}$ & $\mathbf{0.3953}^{\dag\ddag}$ & $\mathbf{0.3830}^{\dag\ddag}$\\
        \midrule
        \midrule
        \multirow{8}{*}{\shortstack[c]{\textbf{Backward}\\\textbf{Setting}}} & \multirow{2}{*}{\textbf{Model}} & \multicolumn{2}{c}{Eng-Fre} & \multicolumn{2}{c}{Eng-Ita} & \multicolumn{2}{c}{Eng-Deu} & \multicolumn{2}{c}{Eng-Spa}\\
        \cmidrule(lr){3-4} \cmidrule(lr){5-6} \cmidrule(lr){7-8} \cmidrule(lr){9-10} 
        & & MAP & P@10 & MAP & P@10 & MAP & P@10 & MAP & P@10 \\
        \cmidrule{2-10}
        & Human Translation & $0.2955$ & $0.3054$ & $0.2629$ & $0.2892$ & $0.2970$ & $0.3060$ & $0.2518$ & $0.2436$\\
        \cmidrule{2-10}
        & SMT & $0.2258$ & $0.2319$ & $0.1883$ & $0.1852$ & $0.2614$ & $0.2424$ & $0.1985$ & $0.2088$\\
        \cmidrule{2-10}
        & \mbertmarco{} & $0.2841^{\dag}$ & $0.2875^{\dag}$ & $0.2635^{\dag}$ & $0.2605^{\dag}$ & $0.3241^{\dag}$ & $0.3246^{\dag}$ & $0.2355^{\dag}$ & $0.2285^{\dag}$\\
        & \matbertplb{} & $0.2807^{\dag}$ & $0.2823^{\dag}$ & $0.2713^{\dag}$ & $0.2771^{\dag}$ & $0.3262^{\dag}$ & $0.3230^{\dag}$ & $0.2389^{\dag}$ & $0.2351^{\dag}$ \\
        & \matbert{} & $\mathbf{0.3002}^{\dag\ddag}$ & $\mathbf{0.3108}^{\dag\ddag}$ & $\mathbf{0.2823}^{\dag\ddag}$ & $\mathbf{0.2846}^{\dag\ddag}$ & $\mathbf{0.3433}^{\dag\ddag}$ & $\mathbf{0.3414}^{\dag\ddag}$ & $\mathbf{0.2558}^{\dag\ddag}$ & $\mathbf{0.2439}^{\dag\ddag}$\\

        \bottomrule
    \end{tabular}
    \end{adjustbox}
\end{table*}

\subsection{Compared Methods.} We compare the proposed model with the methods in  the following
\begin{itemize}[leftmargin=*,noitemsep,topsep=0pt]
    \item \textbf{SMT}: We first use the GIZA++ toolkit~\cite{och2003systematic} to build translation tables from parallel corpora. We select top-10 translations from the translation table for each query term and apply Galago\footnote{\url{ https://www.lemurproject.org/galago.php/}}’s weighted $\#$\textit{combine} operator to form a translated query. Then we use the Galago’s implementation of Okapi BM25~\cite{robertson1995okapi} with default parameters. This setting is taken from~\citet{bonab2020training}, and we call this method statistical machine translation (SMT). It serves as one of our baselines. Moreover, the training data for neural re-ranking models are sampled based on the top 500 retrieved documents by the SMT model.
    
    \item \textbf{\mbertmarco{}}: 
    To create the \mbertmarco{} baseline we begin with the pre-trained checkpoint provided by~\cite{nogueira2019passage}. This checkpoint is a result of fine-tuning the multilingual BERT (mBERT) architecture with MSMARCO passage ranking dataset. We further fine-tune it with training data from a specific CLIR setting. We use the same fine-tuning approach described in section \ref{sec:implementation_details} for this baseline and our proposed model to ensure fair comparison.  
    
    
    
    \item \textbf{\matbertplb{}}: This is a variant of \matbert{}. In order to evaluate the effect of external knowledge on MAT, we replace $M_{tr}$ by an identity matrix so that each token is only paying attention to itself. Therefore, instead of injecting translation knowledge into the model, we design a ``placebo'' attention matrix for MAT. Using \matbertplb{} as a controlled experiment, we are able to evaluate the effect of external knowledge.
\end{itemize} 

In order to provide an empirical upper-bound on retrieval performance, we use human translation of the queries and apply BM25 as the retrieval technique. The human translations of the queries are obtained from the CLEF dataset as they have a common topic ID for the same queries across different languages. 


\section{Experimental Results}

\subsection{Performance on High-resource Languages}
Table~\ref{tab:main-comparison} lists evaluation results on both Forward (top) and Backward (bottom) settings for language pairs with high translation resources.

As a neural re-ranker, \mbertmarco{} significantly improves upon SMT on all language pairs in backward setting and two language pairs on the forward setting while performs on par with SMT for Deu-Eng and Ita-Eng languages. While fine-tuned on English document retrieval dataset, \mbertmarco{} can transfer to cross-lingual task with small amount of fine-tuning data. This agrees with the previous finding by  \citet{pires-etal-2019-multilingual} that mBERT is capable to generalize across languages.


We observed substantial improvements on the retrieval performance when translation knowledge is incorporated into \matbert{}. For all language setting combination in Table~\ref{tab:main-comparison}, \matbert{} performs significantly better than the BERT architecture (\mbertmarco{}) in terms of both MAP and P@10. \matbert{} improves \mbertmarco{} by 8\% on the forward and 7\% on the backward settings in terms of MAP. This comprehensive comparisons with vanilla BERT based ranker demonstrate the effectiveness of the MAT-embedded model.

Replacing $M_{tr}$ by the identity matrix in \matbertplb{}, the translation attention head degenerates to two additional feed-forward layers. \matbertplb{} behaves insignificantly comparing to the vanilla BERT architecture on all languages.
Such results indicate that the performance gain in \matbert{} relies on injecting the external knowledge, not from adding new parameters. When $M_{tr}$ becomes non-informative, the translation attention head is ineffectual.

Comparing \matbert{} with Human Translation, we can see that in forward setting, correct translation with basic retrieval model still lead the neural CLIR model. However, in backward setting, \matbert{} achieves relatively the same as (Eng-Fre, Eng-Spa) or better than (Eng-Ita, Eng-Deu) Human Translation. We hypothesize that in the backward setting, translation tables provide higher quality translations 
which enable better semantic matching between query and document tokens.

\begin{figure*}[t]
    \captionsetup[subfigure]{aboveskip=2pt,font=footnotesize,labelfont=footnotesize}
    \begin{subfigure}[t]{0.45\textwidth}
        \centering
        \includegraphics[width=0.8\linewidth]{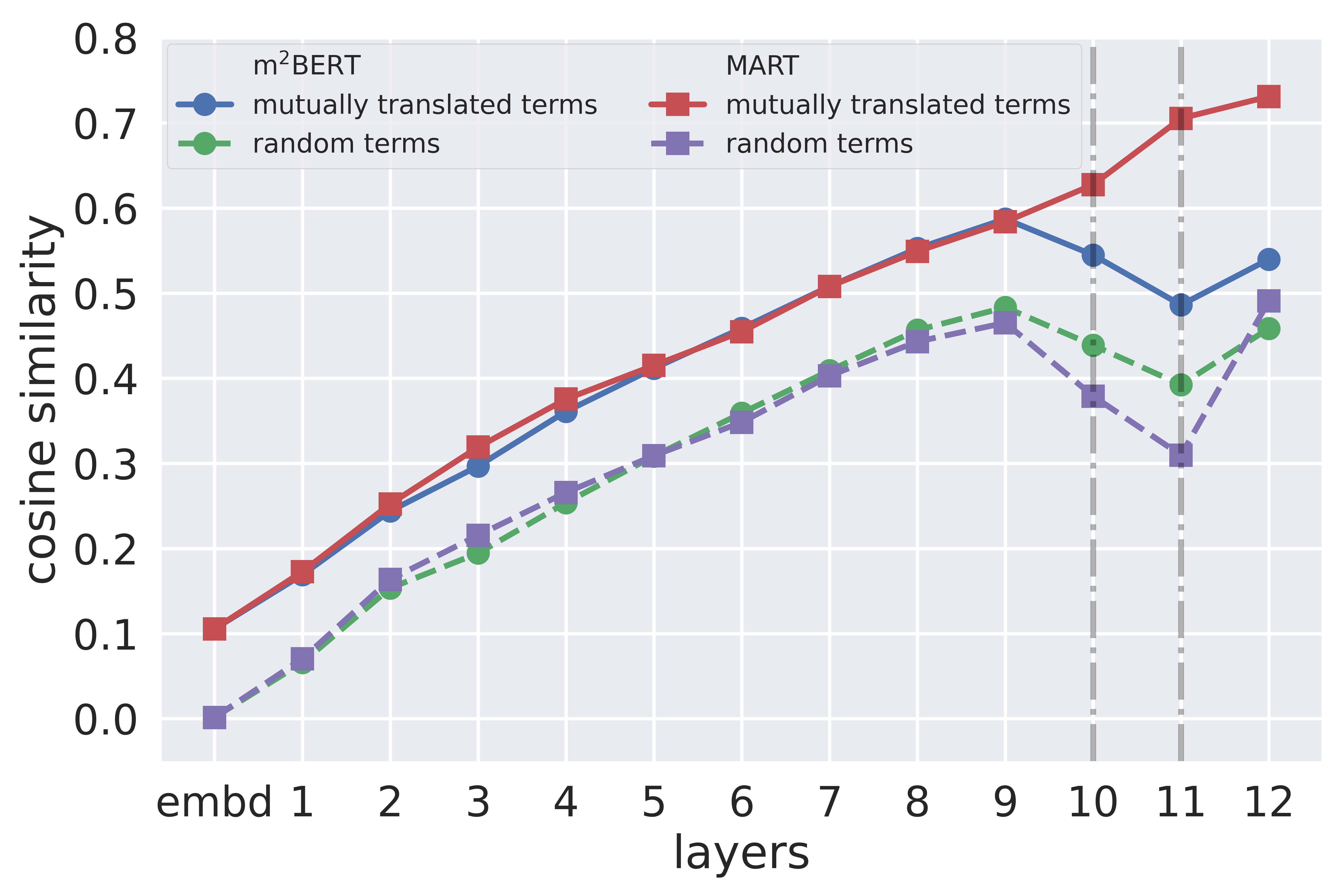}
        \vspace{-0.1cm}
        \caption{Deu-Eng}
    \end{subfigure}
    \begin{subfigure}[t]{0.45\textwidth}
        \centering
        \includegraphics[width=0.8\linewidth]{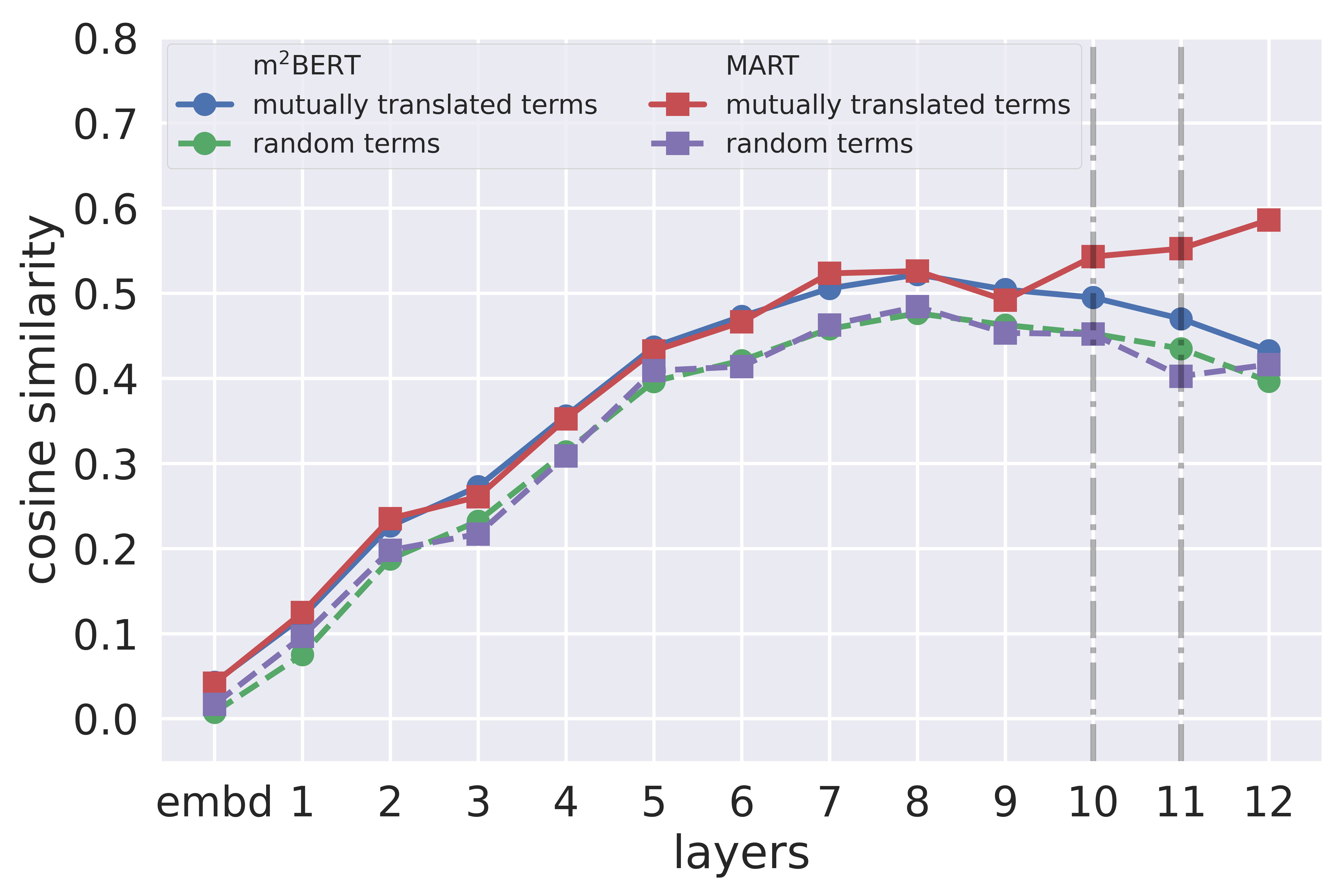}
        \vspace{-0.1cm}
        \caption{Swa-Eng}
    \end{subfigure}
\vspace{-0.3cm}
\caption{The comparison of \matbert{} to \mbertmarco{} on layer-wise token representations.}
\vspace{-0.1cm}
\label{fig:layer-analysis}
\end{figure*}

\subsection{Performance on Low-resource Languages}
The evaluation results for two language pairs with limited translation resources on the forward setting are shown in Table~\ref{tab:low-resource}. We make several observations.
First, \mbertmarco{} mostly under-performs SMT for both Somali and Swahili languages.
Note that Somali is not included in the pre-training of mBERT. Even if Swahili is included, there is only a small number of Swahili sentences in the pre-training data.
The low performance of \mbertmarco{} on low-resource language pairs demonstrates that absence or inadequate pre-training data on a particular language leads to poor performance on target tasks involving those languages. 

On the other hand, the \matbert{} model achieves the highest MAP performance for both Somali and Swahili languages. The consistent and significant improvements in terms of MAP over compared methods make \matbert{} the best model in our experiments. 
Due to the lack of pre-training data, the translation gap is more critical in low-resource language pairs. The performance of \matbert{} for Somali and Swahili languages proves that leveraging the external translation knowledge can help to bridge the translation gap.
Moreover, the experiments with the placebo setting, similar to those for the high-resource languages, have shown no significance in performance compared to \mbertmarco{}. These results strengthen the conclusion that the translation attention matrix is the key component of MAT.

Human Translation leads neural ranking models by a large margin in CLIR tasks involving low-resource languages. This is expected because, with less sentence-level parallel data, the CLIR models often suffer from low  quality of translations.

\begin{table}[t]
    \centering
    \captionsetup{width=\linewidth}
    \caption{Model performance for low-resource languages on Forward setting. The highest value for each column is marked with bold text. Statistically significant improvements are marked by $\dag$ (over SMT) and $\ddag$ (over \mbertmarco{}).}
    \label{tab:low-resource}
    \begin{adjustbox}{width=0.43\textwidth}
    \aboverulesep=0ex
    \belowrulesep=0ex
    \renewcommand{\arraystretch}{1.2}
    \begin{tabular}{lcccc}
        \toprule
        \multirow{2}{*}{\textbf{Model}} & \multicolumn{2}{c}{Som-Eng} & \multicolumn{2}{c}{Swa-Eng}\\
        \cmidrule(lr){2-3} \cmidrule(lr){4-5}
        & MAP & P@10 & MAP & P@10\\
        \midrule
        Human Translation & $0.4563$ & $ 0.3940$ & $0.4563$ & $ 0.3940$ \\
        \hline
        SMT & $0.1948$ & $0.1865$ & $0.2184$ & $\mathbf{0.2152}$\\
        \hline
        \mbertmarco{} & $0.1986$ & $0.1772$ & $0.2055$ & $0.2089$\\
        \matbertplb{} & $0.2049$ & $0.1972^{\dag\ddag}$ & $0.2130$ & $0.2106$\\
        \matbert{} & $\mathbf{0.2207}^{\dag\ddag}$ & $\mathbf{0.2135}^{\dag\ddag}$ & $\mathbf{0.2348}^{\dag\ddag}$ & $0.2151$\\
        \bottomrule
    \end{tabular}
    \end{adjustbox}
\end{table}

\begin{figure*}[t]
    \captionsetup[subfigure]{aboveskip=2pt,font=footnotesize,labelfont=footnotesize}
    \begin{subfigure}[t]{0.33\textwidth}
        \centering
        \includegraphics[width=\linewidth]{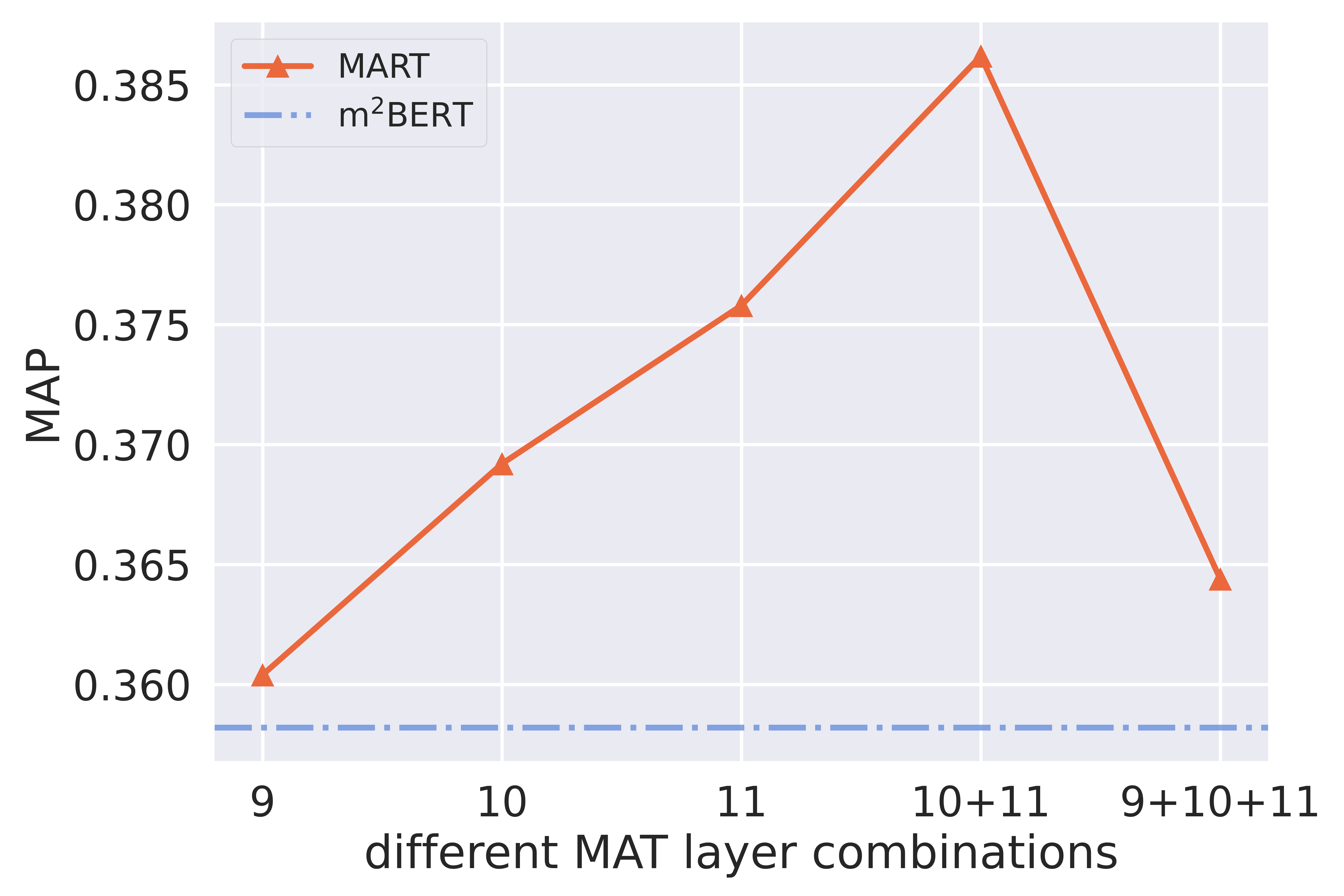}
        \caption{Deu-Eng}
        \label{fig:layer-deu-eng}
    \end{subfigure}\hspace{-0.3em}%
    \begin{subfigure}[t]{0.33\textwidth}
        \centering
        \includegraphics[width=\linewidth]{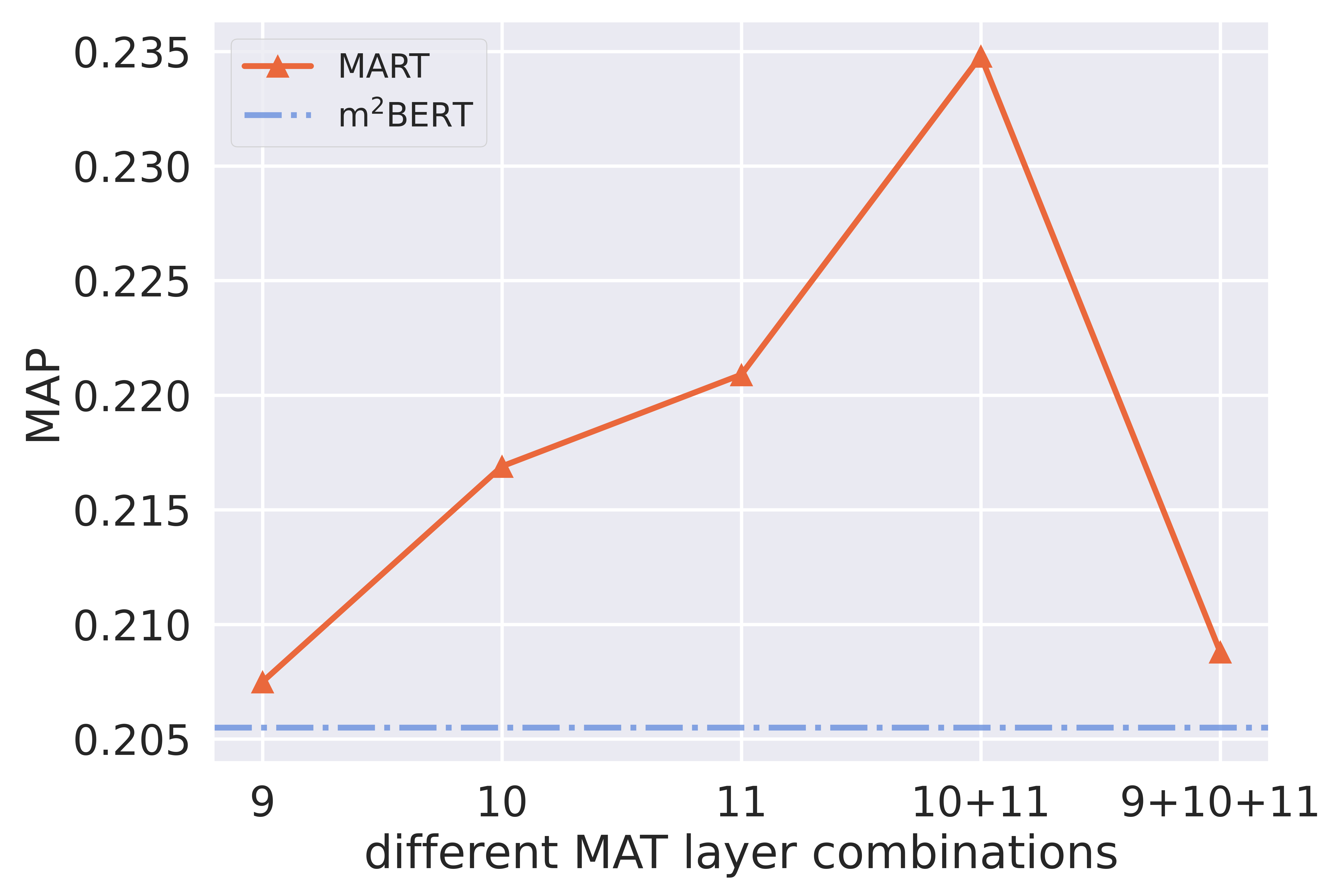}
        \caption{Swa-Eng}
        \label{fig:layer-swa-eng}
    \end{subfigure}\hspace{-0.5em}
    \begin{subfigure}[t]{0.33\textwidth}
        \centering
        \includegraphics[width=\linewidth]{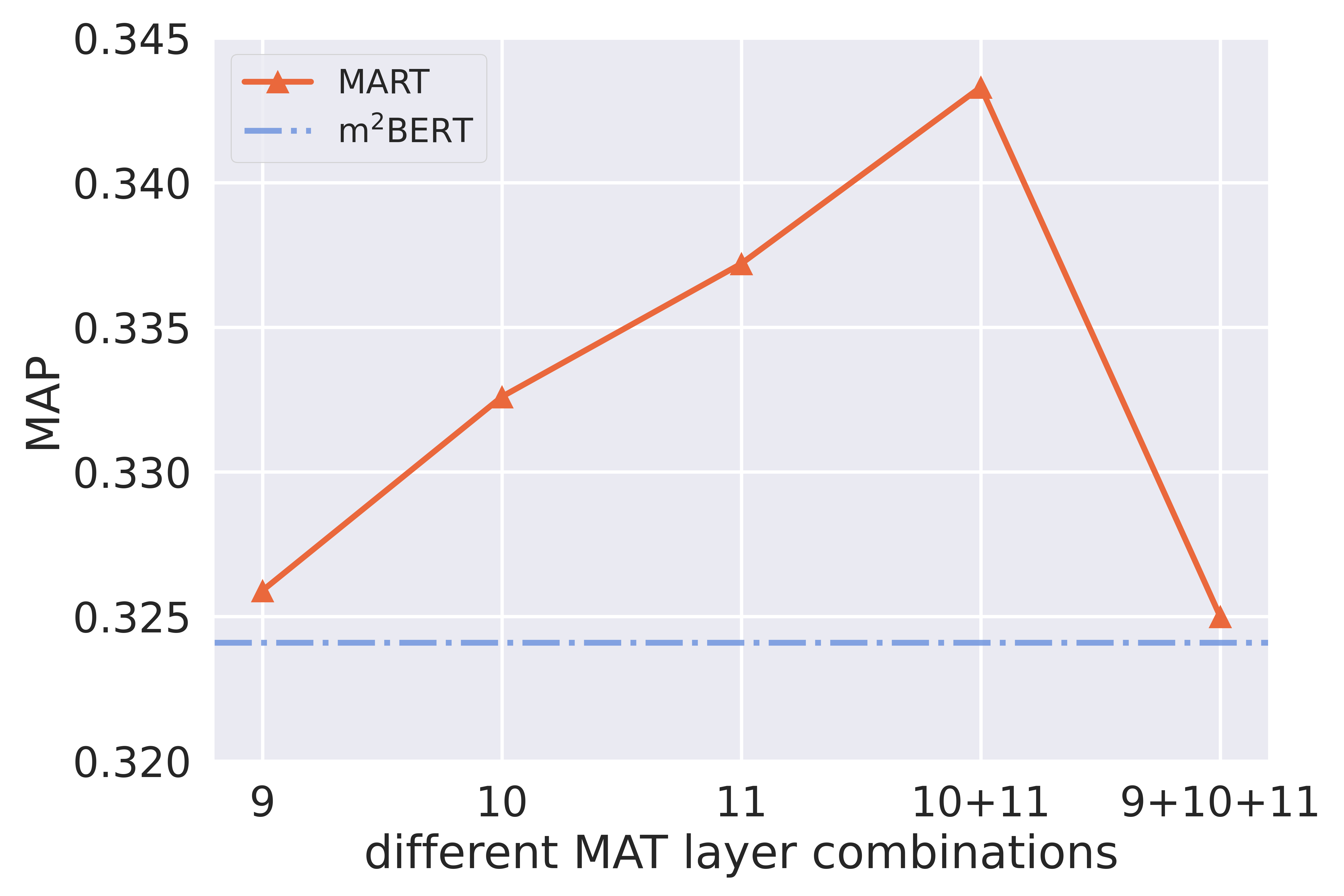}
        \caption{Eng-Deu}
        \label{fig:layer-eng-deu}
    \end{subfigure}
\vspace{-0.3cm}
\caption{The performance comparison of different \matbert{} model architectures.}
\vspace{-0.1cm}
\label{fig:model-analysis}
\end{figure*}

\subsection{Representation Analysis}
To study the influence of MAT on the translation gap in neural CLIR, we compare the token representation from each layer between \mbertmarco{} and \matbert{}.
Specifically, both models are fine-tuned on Deu-Eng and Swa-Eng training data. Figure~\ref{fig:layer-analysis} shows the distances between contextualized token representations in two model architectures where x-axis represents layers from low to high and y-axis is the cosine similarity. We focus on two types of word pairs (one from query and another from document) in an input sequence: (i)~Mutually translated words, where all pairs of words that are translations to each other according to the external translation knowledge are selected; and (ii)~Random non-translated words, where we randomly sample 10 pairs of words which are not translations of each other. We compute the average cosine similarity of the token representations at each layer for all selected word pairs in the test data of Deu-Eng (high-resource) and Swa-Eng (low-resource).

From the diagrams in Figure~\ref{fig:layer-analysis}, we can see that in general, the similarity of token representations increases as the layer gets higher. Also, the mutually translated words always have smaller cosine distances than non-translated words. The closer lines between two types of word pairs in Swa-Eng prove that the translation gap is more critical in resource-lean languages.
We can also see that in 10\textsuperscript{th} and 11\textsuperscript{th} layers,  the  similarity of two types of words in \mbertmarco{} drops for both language pairs. According to the previous analysis~\cite{pires-etal-2019-multilingual}, one hypothesis for such drop is that before fine-tuning on MS MARCO dataset, mBERT was pre-trained on surrounding contexts for language modeling, it needs more contextual information to correctly predict the missing words. Therefore, mBERT favors text sequence pairs that are closer in their semantic meanings. Such models trained on surrounding context   are not as effective for ad-hoc document ranking with respect to
keyword queries~\cite{qiao2019understanding}. 

\matbert{} shows the same behavior as \mbertmarco{} up to the MAT layers. The representations of mutually translated words in MAT layers become similar to each other in terms of cosine distance. This matches the design purpose of MAT. 
Meanwhile, because MAT keeps the native multi-head attention from BERT layer, the similarity of non-translations still drops in MAT layers.
The increased similarity on mutually translated words and decreased similarity on non-translated words demonstrate that model is bridging translation gap with the help of external knowledge.

\begin{table}[t]
    \centering
    \captionsetup{width=\linewidth}
    \caption{\matbert{} performance for different external knowledge. The highest value for each column is marked with bold text. ``\textbf{--}'' if language is not supported.}
    \label{tab:translation-resource}
    \begin{adjustbox}{width=0.48\textwidth}
    \aboverulesep=0ex
    \belowrulesep=0ex
    \renewcommand{\arraystretch}{1.2}
    \begin{tabular}{lcccc|cc}
        \toprule
        \multirow{3}{*}{\shortstack[c]{\textbf{External}\\\textbf{Knowledge}}} & \multicolumn{4}{c|}{Forward} & \multicolumn{2}{c}{Backward} \\
        \cmidrule(lr){2-7}
        & \multicolumn{2}{c}{Deu-Eng} & \multicolumn{2}{c|}{Swa-Eng} &  \multicolumn{2}{c}{Eng-Deu}\\
        \cmidrule(lr){2-3} \cmidrule(lr){4-5} \cmidrule(lr){6-7}
        & MAP & P@10 & MAP & P@10 & MAP & P@10 \\
        \midrule
        Parallel Corpus & $\mathbf{0.3862}$ & $\mathbf{0.3770}$ & $\mathbf{0.2348}$ & $\mathbf{0.2151}$ & $\mathbf{0.3433}$ & $\mathbf{0.3414}$ \\
        Panlex & $0.3713$ & $0.3612$ & $0.2265$ & $0.2073$ & $0.3326$ & $0.3360$\\
        MUSE & $0.3693$ & $0.3580$ & \textbf{--} & \textbf{--} & $0.3335$ & $0.3348$\\
        \bottomrule
    \end{tabular}
    \end{adjustbox}
\end{table}

\subsection{Effect of Translation Resources}

From the previous results, we have seen that the translation attention matrix is critical to the success of MAT. As a knowledge injection model, it is palpable that the quality of the knowledge affects the model performance. In this experiment, we study the effect of different sources of external knowledge on the \matbert{}. Besides the translation table built from parallel data, we use two different translation knowledge for $M_{tr}$ generation: Panlex dictionary~\cite{kamholz2014panlex} and multilingual word embedding (MUSE~\cite{conneau2017word}). To obtain translation probability for a single word in Panlex, we uniformly distribute weights to all possible translations. And in MUSE, we use the 5 nearest neighbors of a word in the target languages as its potential translations and assign translation probability based on their normalized cosine similarity. In order to cover different languages and retrieval settings, we select Deu-Eng (high-resource) and Swa-Eng (low-resource) from forward setting and Eng-Deu from backward setting for this experiment.

Table~\ref{tab:translation-resource} shows the results of all compared translation knowledge. We observe a performance drop on both alternative knowledge resources.
For Panlex, although the translations are more precise than those in a translation table, they do not provide a broad coverage of words. 
Multilingual word embeddings are learnt from the contexts of words, not their translations. Therefore, given a word, the embeddings of semantically similar words are often closer than those of its translations to the embedding of a word~\cite{bonab2020training}. Thus, using multilingual word embeddings, the problem of the translation gap will not be completely resolved.


\subsection{Ablation Study on Model Architecture}

In this section, we empirically study the effects of different numbers and positions of MAT layers in a \matbert{} model. We further train and evaluate the \matbert{} with various combinations of MAT layers. It is worth mentioning that given the number of layers in BERT architecture, there exist exponential number of possible combinations. We only explore several representative models. Leaving the last layer as the output layer, we still focus on the higher transformer layers of BERT architecture. For models with a single MAT layer, we investigate \matbert{} with MAT embedded at 9\textsuperscript{th}, 10\textsuperscript{th}, or 11\textsuperscript{th} layer. For double MAT layers, we use the previous results from MAT at 10\textsuperscript{th} and 11\textsuperscript{th} layers. We also consider an architecture with three MAT layers where 9\textsuperscript{th}, 10\textsuperscript{th} and 11\textsuperscript{th} layers in BERT are all replaced by the MAT layer.

Figure~\ref{fig:model-analysis} shows the performance of different \matbert{} model architectures on Deu-Eng, Swa-Eng and Eng-Deu. We can see that all model variants have the similar pattern across three selected CLIR tasks. Because higher BERT layers are more sensitive to fine-tuning~\cite{zhao-bethard-2020-berts} and their hidden representations capture complex semantic information~\cite{tenney-etal-2019-bert}, the retrieval performance for the single MAT layer increases from MAT at the 9\textsuperscript{th} layer to MAT at the 11\textsuperscript{th} layer. The double MAT layer can further boost performance from the single-layer approach. We also can see that models get less improved when 9\textsuperscript{th} in replaced by MAT. 
We hypothesize that the token representations after the 8\textsuperscript{th} layer (the input of the 9\textsuperscript{th} layer) do not contain enough semantic information~\cite{tenney-etal-2019-bert} so it is too early to apply the translation attentions. 


\section{Conclusion}  \label{sec:conclusion}

In this paper, we propose a novel Mixed Attention Transformer (MAT) network to leverage external translation knowledge for cross-lingual information retrieval tasks.

First, we build attention matrix for mutually translated words between query and document based on the translation resource. Then using the attention matrix, we design a new translation attention head and show that it is able to reduce the cosine distance between hidden representations of mutually translated words.
Finally, the complete architecture of MAT is a combination of multi-head attention and translation attention head with shared feed-forward networks.
As a layer component, we further design a sandwich-like architecture to embed MAT into the Transformer model.
Our comprehensive experimental results demonstrate the effectiveness of external knowledge and the significant improvement of MAT-embedded neural model on CLIR task.

For future work, we are particularly interested in fine-tuning \matbert{} on a large CLIR dataset with a mix of cross-language settings to learn a language-agnostic neural ranking model. 
We also plan to apply MAT to other retrieval tasks, e.g., event retrieval, by incorporating information other than translation knowledge.
\begin{acks}
This work was supported in part by the Center for Intelligent Information Retrieval, in part under USC (University of Southern California) subcontract no. 124338456 under IARPA prime contract no. 2019-19051600007., and in part by the Office of the Director of National Intelligence (ODNI), Intelligence Advanced Research Projects Activity (IARPA) via AFRL contact \#FA8650-17-C-9116 under subcontract \#94671240 from the University of Southern California. The views and conclusions contained herein are those of the authors and should not be interpreted as necessarily representing the official policies or endorsements, either expressed or implied, of the ODNI, IARPA, or the U.S. Government. The U.S. Government is authorized to reproduce and distribute reprints for Governmental purposes notwithstanding any copyright annotation thereon.
\end{acks}

\bibliographystyle{ACM-Reference-Format}
\bibliography{reference}


\begin{thebibliography}{53}


\ifx \showCODEN    \undefined \def \showCODEN     #1{\unskip}     \fi
\ifx \showDOI      \undefined \def \showDOI       #1{#1}\fi
\ifx \showISBNx    \undefined \def \showISBNx     #1{\unskip}     \fi
\ifx \showISBNxiii \undefined \def \showISBNxiii  #1{\unskip}     \fi
\ifx \showISSN     \undefined \def \showISSN      #1{\unskip}     \fi
\ifx \showLCCN     \undefined \def \showLCCN      #1{\unskip}     \fi
\ifx \shownote     \undefined \def \shownote      #1{#1}          \fi
\ifx \showarticletitle \undefined \def \showarticletitle #1{#1}   \fi
\ifx \showURL      \undefined \def \showURL       {\relax}        \fi
\providecommand\bibfield[2]{#2}
\providecommand\bibinfo[2]{#2}
\providecommand\natexlab[1]{#1}
\providecommand\showeprint[2][]{arXiv:#2}

\bibitem[\protect\citeauthoryear{Bonab, Allan, and Sitaraman}{Bonab
  et~al\mbox{.}}{2019}]%
        {10.1145/3341981.3344236}
\bibfield{author}{\bibinfo{person}{Hamed Bonab}, \bibinfo{person}{James Allan},
  {and} \bibinfo{person}{Ramesh Sitaraman}.} \bibinfo{year}{2019}\natexlab{}.
\newblock \showarticletitle{Simulating CLIR Translation Resource Scarcity Using
  High-Resource Languages}. In \bibinfo{booktitle}{\emph{Proceedings of the
  2019 ACM SIGIR International Conference on Theory of Information Retrieval}}
  (Santa Clara, CA, USA) \emph{(\bibinfo{series}{ICTIR '19})}.
  \bibinfo{publisher}{Association for Computing Machinery},
  \bibinfo{address}{New York, NY, USA}, \bibinfo{pages}{129–136}.
\newblock
\showISBNx{9781450368810}
\urldef\tempurl%
\url{https://doi.org/10.1145/3341981.3344236}
\showDOI{\tempurl}


\bibitem[\protect\citeauthoryear{Bonab, Sarwar, and Allan}{Bonab
  et~al\mbox{.}}{2020}]%
        {bonab2020training}
\bibfield{author}{\bibinfo{person}{Hamed Bonab},
  \bibinfo{person}{Sheikh~Muhammad Sarwar}, {and} \bibinfo{person}{James
  Allan}.} \bibinfo{year}{2020}\natexlab{}.
\newblock \showarticletitle{Training Effective Neural CLIR by Bridging the
  Translation Gap}. In \bibinfo{booktitle}{\emph{Proceedings of the 43rd
  International ACM SIGIR Conference on Research and Development in Information
  Retrieval}}. \bibinfo{pages}{9--18}.
\newblock


\bibitem[\protect\citeauthoryear{Braschler}{Braschler}{2001}]%
        {10.1007/3-540-44645-1_9}
\bibfield{author}{\bibinfo{person}{Martin Braschler}.}
  \bibinfo{year}{2001}\natexlab{}.
\newblock \showarticletitle{CLEF 2000 --- Overview of Results}. In
  \bibinfo{booktitle}{\emph{Cross-Language Information Retrieval and
  Evaluation}}, \bibfield{editor}{\bibinfo{person}{Carol Peters}} (Ed.).
  \bibinfo{publisher}{Springer Berlin Heidelberg}, \bibinfo{address}{Berlin,
  Heidelberg}, \bibinfo{pages}{89--101}.
\newblock
\showISBNx{978-3-540-44645-3}


\bibitem[\protect\citeauthoryear{Braschler}{Braschler}{2002a}]%
        {10.1007/3-540-45691-0_2}
\bibfield{author}{\bibinfo{person}{Martin Braschler}.}
  \bibinfo{year}{2002}\natexlab{a}.
\newblock \showarticletitle{CLEF 2001 --- Overview of Results}. In
  \bibinfo{booktitle}{\emph{Evaluation of Cross-Language Information Retrieval
  Systems}}, \bibfield{editor}{\bibinfo{person}{Carol Peters},
  \bibinfo{person}{Martin Braschler}, \bibinfo{person}{Julio Gonzalo}, {and}
  \bibinfo{person}{Michael Kluck}} (Eds.). \bibinfo{publisher}{Springer Berlin
  Heidelberg}, \bibinfo{address}{Berlin, Heidelberg}, \bibinfo{pages}{9--26}.
\newblock
\showISBNx{978-3-540-45691-9}


\bibitem[\protect\citeauthoryear{Braschler}{Braschler}{2002b}]%
        {braschler2002clef}
\bibfield{author}{\bibinfo{person}{Martin Braschler}.}
  \bibinfo{year}{2002}\natexlab{b}.
\newblock \showarticletitle{CLEF 2002—Overview of results}. In
  \bibinfo{booktitle}{\emph{Workshop of the Cross-Language Evaluation Forum for
  European Languages}}. Springer, \bibinfo{pages}{9--27}.
\newblock


\bibitem[\protect\citeauthoryear{Braschler}{Braschler}{2003}]%
        {braschler2003clef}
\bibfield{author}{\bibinfo{person}{Martin Braschler}.}
  \bibinfo{year}{2003}\natexlab{}.
\newblock \showarticletitle{CLEF 2003--Overview of results}. In
  \bibinfo{booktitle}{\emph{Workshop of the Cross-Language Evaluation Forum for
  European Languages}}. Springer, \bibinfo{pages}{44--63}.
\newblock


\bibitem[\protect\citeauthoryear{Conneau, Khandelwal, Goyal, Chaudhary, Wenzek,
  Guzm{\'a}n, Grave, Ott, Zettlemoyer, and Stoyanov}{Conneau
  et~al\mbox{.}}{2020a}]%
        {conneau2020unsupervised}
\bibfield{author}{\bibinfo{person}{Alexis Conneau}, \bibinfo{person}{Kartikay
  Khandelwal}, \bibinfo{person}{Naman Goyal}, \bibinfo{person}{Vishrav
  Chaudhary}, \bibinfo{person}{Guillaume Wenzek}, \bibinfo{person}{Francisco
  Guzm{\'a}n}, \bibinfo{person}{{\'E}douard Grave}, \bibinfo{person}{Myle Ott},
  \bibinfo{person}{Luke Zettlemoyer}, {and} \bibinfo{person}{Veselin
  Stoyanov}.} \bibinfo{year}{2020}\natexlab{a}.
\newblock \showarticletitle{Unsupervised Cross-lingual Representation Learning
  at Scale}. In \bibinfo{booktitle}{\emph{Proceedings of the 58th Annual
  Meeting of the Association for Computational Linguistics}}.
  \bibinfo{pages}{8440--8451}.
\newblock


\bibitem[\protect\citeauthoryear{Conneau, Khandelwal, Goyal, Chaudhary, Wenzek,
  Guzm{\'a}n, Grave, Ott, Zettlemoyer, and Stoyanov}{Conneau
  et~al\mbox{.}}{2020b}]%
        {conneau-etal-2020-unsupervised}
\bibfield{author}{\bibinfo{person}{Alexis Conneau}, \bibinfo{person}{Kartikay
  Khandelwal}, \bibinfo{person}{Naman Goyal}, \bibinfo{person}{Vishrav
  Chaudhary}, \bibinfo{person}{Guillaume Wenzek}, \bibinfo{person}{Francisco
  Guzm{\'a}n}, \bibinfo{person}{Edouard Grave}, \bibinfo{person}{Myle Ott},
  \bibinfo{person}{Luke Zettlemoyer}, {and} \bibinfo{person}{Veselin
  Stoyanov}.} \bibinfo{year}{2020}\natexlab{b}.
\newblock \showarticletitle{Unsupervised Cross-lingual Representation Learning
  at Scale}. In \bibinfo{booktitle}{\emph{Proceedings of the 58th Annual
  Meeting of the Association for Computational Linguistics}}.
  \bibinfo{publisher}{Association for Computational Linguistics},
  \bibinfo{address}{Online}, \bibinfo{pages}{8440--8451}.
\newblock
\urldef\tempurl%
\url{https://doi.org/10.18653/v1/2020.acl-main.747}
\showDOI{\tempurl}


\bibitem[\protect\citeauthoryear{Conneau, Lample, Ranzato, Denoyer, and
  J{\'e}gou}{Conneau et~al\mbox{.}}{2017}]%
        {conneau2017word}
\bibfield{author}{\bibinfo{person}{Alexis Conneau}, \bibinfo{person}{Guillaume
  Lample}, \bibinfo{person}{Marc'Aurelio Ranzato}, \bibinfo{person}{Ludovic
  Denoyer}, {and} \bibinfo{person}{Herv{\'e} J{\'e}gou}.}
  \bibinfo{year}{2017}\natexlab{}.
\newblock \showarticletitle{Word translation without parallel data}.
\newblock \bibinfo{journal}{\emph{arXiv preprint arXiv:1710.04087}}
  (\bibinfo{year}{2017}).
\newblock


\bibitem[\protect\citeauthoryear{Correia, Niculae, and Martins}{Correia
  et~al\mbox{.}}{2019}]%
        {correia-etal-2019-adaptively}
\bibfield{author}{\bibinfo{person}{Gon{\c{c}}alo~M. Correia},
  \bibinfo{person}{Vlad Niculae}, {and} \bibinfo{person}{Andr{\'e} F.~T.
  Martins}.} \bibinfo{year}{2019}\natexlab{}.
\newblock \showarticletitle{Adaptively Sparse Transformers}. In
  \bibinfo{booktitle}{\emph{Proceedings of the 2019 Conference on Empirical
  Methods in Natural Language Processing and the 9th International Joint
  Conference on Natural Language Processing (EMNLP-IJCNLP)}}.
  \bibinfo{publisher}{Association for Computational Linguistics},
  \bibinfo{address}{Hong Kong, China}, \bibinfo{pages}{2174--2184}.
\newblock
\urldef\tempurl%
\url{https://doi.org/10.18653/v1/D19-1223}
\showDOI{\tempurl}


\bibitem[\protect\citeauthoryear{Dehghani, Zamani, Severyn, Kamps, and
  Croft}{Dehghani et~al\mbox{.}}{2017}]%
        {dehghani2017neural}
\bibfield{author}{\bibinfo{person}{Mostafa Dehghani}, \bibinfo{person}{Hamed
  Zamani}, \bibinfo{person}{Aliaksei Severyn}, \bibinfo{person}{Jaap Kamps},
  {and} \bibinfo{person}{W~Bruce Croft}.} \bibinfo{year}{2017}\natexlab{}.
\newblock \showarticletitle{Neural ranking models with weak supervision}. In
  \bibinfo{booktitle}{\emph{Proceedings of the 40th International ACM SIGIR
  Conference on Research and Development in Information Retrieval}}.
  \bibinfo{pages}{65--74}.
\newblock


\bibitem[\protect\citeauthoryear{Devlin, Chang, Lee, and Toutanova}{Devlin
  et~al\mbox{.}}{2019}]%
        {devlin-etal-2019-bert}
\bibfield{author}{\bibinfo{person}{Jacob Devlin}, \bibinfo{person}{Ming-Wei
  Chang}, \bibinfo{person}{Kenton Lee}, {and} \bibinfo{person}{Kristina
  Toutanova}.} \bibinfo{year}{2019}\natexlab{}.
\newblock \showarticletitle{{BERT}: Pre-training of Deep Bidirectional
  Transformers for Language Understanding}. In
  \bibinfo{booktitle}{\emph{Proceedings of the 2019 Conference of the North
  {A}merican Chapter of the Association for Computational Linguistics: Human
  Language Technologies, Volume 1 (Long and Short Papers)}}.
  \bibinfo{publisher}{Association for Computational Linguistics},
  \bibinfo{address}{Minneapolis, Minnesota}, \bibinfo{pages}{4171--4186}.
\newblock
\urldef\tempurl%
\url{https://doi.org/10.18653/v1/N19-1423}
\showDOI{\tempurl}


\bibitem[\protect\citeauthoryear{Guo, Fan, Ai, and Croft}{Guo
  et~al\mbox{.}}{2016}]%
        {Guo_2016}
\bibfield{author}{\bibinfo{person}{Jiafeng Guo}, \bibinfo{person}{Yixing Fan},
  \bibinfo{person}{Qingyao Ai}, {and} \bibinfo{person}{W.~Bruce Croft}.}
  \bibinfo{year}{2016}\natexlab{}.
\newblock \showarticletitle{A Deep Relevance Matching Model for Ad-hoc
  Retrieval}.
\newblock \bibinfo{journal}{\emph{Proceedings of the 25th ACM International on
  Conference on Information and Knowledge Management}} (\bibinfo{date}{Oct}
  \bibinfo{year}{2016}).
\newblock
\showISBNx{9781450340731}
\urldef\tempurl%
\url{https://doi.org/10.1145/2983323.2983769}
\showDOI{\tempurl}


\bibitem[\protect\citeauthoryear{He, Zhou, Xiao, Jiang, Liu, Yuan, and Xu}{He
  et~al\mbox{.}}{2020}]%
        {He2020IntegratingGC}
\bibfield{author}{\bibinfo{person}{Bin He}, \bibinfo{person}{Di Zhou},
  \bibinfo{person}{JingHui Xiao}, \bibinfo{person}{X. Jiang},
  \bibinfo{person}{Qun Liu}, \bibinfo{person}{Nicholas~Jing Yuan}, {and}
  \bibinfo{person}{T. Xu}.} \bibinfo{year}{2020}\natexlab{}.
\newblock \showarticletitle{Integrating Graph Contextualized Knowledge into
  Pre-trained Language Models}.
\newblock \bibinfo{journal}{\emph{ArXiv}}  \bibinfo{volume}{abs/1912.00147}
  (\bibinfo{year}{2020}).
\newblock


\bibitem[\protect\citeauthoryear{Jiang, El-Jaroudi, Hartmann, Karakos, and
  Zhao}{Jiang et~al\mbox{.}}{2020}]%
        {jiang-etal-2020-cross}
\bibfield{author}{\bibinfo{person}{Zhuolin Jiang}, \bibinfo{person}{Amro
  El-Jaroudi}, \bibinfo{person}{William Hartmann}, \bibinfo{person}{Damianos
  Karakos}, {and} \bibinfo{person}{Lingjun Zhao}.}
  \bibinfo{year}{2020}\natexlab{}.
\newblock \showarticletitle{Cross-lingual Information Retrieval with {BERT}}.
  In \bibinfo{booktitle}{\emph{Proceedings of the workshop on Cross-Language
  Search and Summarization of Text and Speech (CLSSTS2020)}}.
  \bibinfo{publisher}{European Language Resources Association},
  \bibinfo{address}{Marseille, France}, \bibinfo{pages}{26--31}.
\newblock
\showISBNx{979-10-95546-55-9}
\urldef\tempurl%
\url{https://www.aclweb.org/anthology/2020.clssts-1.5}
\showURL{%
\tempurl}


\bibitem[\protect\citeauthoryear{Kamholz, Pool, and Colowick}{Kamholz
  et~al\mbox{.}}{2014}]%
        {kamholz2014panlex}
\bibfield{author}{\bibinfo{person}{David Kamholz}, \bibinfo{person}{Jonathan
  Pool}, {and} \bibinfo{person}{Susan~M Colowick}.}
  \bibinfo{year}{2014}\natexlab{}.
\newblock \showarticletitle{PanLex: Building a Resource for Panlingual Lexical
  Translation.}. In \bibinfo{booktitle}{\emph{LREC}}.
  \bibinfo{pages}{3145--3150}.
\newblock


\bibitem[\protect\citeauthoryear{Kingma and Ba}{Kingma and Ba}{2015}]%
        {kingma2014adam}
\bibfield{author}{\bibinfo{person}{Diederik~P Kingma} {and}
  \bibinfo{person}{Jimmy Ba}.} \bibinfo{year}{2015}\natexlab{}.
\newblock \showarticletitle{Adam: A method for stochastic optimization}.
\newblock  (\bibinfo{year}{2015}).
\newblock


\bibitem[\protect\citeauthoryear{Koehn}{Koehn}{2005}]%
        {koehn2005europarl}
\bibfield{author}{\bibinfo{person}{Philipp Koehn}.}
  \bibinfo{year}{2005}\natexlab{}.
\newblock \showarticletitle{Europarl: A parallel corpus for statistical machine
  translation}. In \bibinfo{booktitle}{\emph{MT summit}},
  Vol.~\bibinfo{volume}{5}. Citeseer, \bibinfo{pages}{79--86}.
\newblock


\bibitem[\protect\citeauthoryear{Lauscher, Vuli'c, Ponti, Korhonen, and
  Glavavs}{Lauscher et~al\mbox{.}}{2020}]%
        {Lauscher2020SpecializingUP}
\bibfield{author}{\bibinfo{person}{Anne Lauscher}, \bibinfo{person}{Ivan
  Vuli'c}, \bibinfo{person}{E. Ponti}, \bibinfo{person}{A. Korhonen}, {and}
  \bibinfo{person}{Goran Glavavs}.} \bibinfo{year}{2020}\natexlab{}.
\newblock \showarticletitle{Specializing Unsupervised Pretraining Models for
  Word-Level Semantic Similarity}. In \bibinfo{booktitle}{\emph{COLING}}.
\newblock


\bibitem[\protect\citeauthoryear{Levine, Lenz, Dagan, Ram, Padnos, Sharir,
  Shalev-Shwartz, Shashua, and Shoham}{Levine et~al\mbox{.}}{2020}]%
        {Levine2020SenseBERTDS}
\bibfield{author}{\bibinfo{person}{Yoav Levine}, \bibinfo{person}{Barak Lenz},
  \bibinfo{person}{Or Dagan}, \bibinfo{person}{Ori Ram}, \bibinfo{person}{Dan
  Padnos}, \bibinfo{person}{Or Sharir}, \bibinfo{person}{S. Shalev-Shwartz},
  \bibinfo{person}{A. Shashua}, {and} \bibinfo{person}{Y. Shoham}.}
  \bibinfo{year}{2020}\natexlab{}.
\newblock \showarticletitle{SenseBERT: Driving Some Sense into BERT}.
\newblock \bibinfo{journal}{\emph{ArXiv}}  \bibinfo{volume}{abs/1908.05646}
  (\bibinfo{year}{2020}).
\newblock


\bibitem[\protect\citeauthoryear{Li and Cheng}{Li and Cheng}{2018}]%
        {li2018learning}
\bibfield{author}{\bibinfo{person}{Bo Li} {and} \bibinfo{person}{Ping Cheng}.}
  \bibinfo{year}{2018}\natexlab{}.
\newblock \showarticletitle{Learning neural representation for clir with
  adversarial framework}. In \bibinfo{booktitle}{\emph{Proceedings of the 2018
  Conference on Empirical Methods in Natural Language Processing}}.
  \bibinfo{pages}{1861--1870}.
\newblock


\bibitem[\protect\citeauthoryear{Litschko, Glava{\v{s}}, Ponzetto, and
  Vuli{\'c}}{Litschko et~al\mbox{.}}{2018}]%
        {litschko2018unsupervised}
\bibfield{author}{\bibinfo{person}{Robert Litschko}, \bibinfo{person}{Goran
  Glava{\v{s}}}, \bibinfo{person}{Simone~Paolo Ponzetto}, {and}
  \bibinfo{person}{Ivan Vuli{\'c}}.} \bibinfo{year}{2018}\natexlab{}.
\newblock \showarticletitle{Unsupervised cross-lingual information retrieval
  using monolingual data only}. In \bibinfo{booktitle}{\emph{The 41st
  International ACM SIGIR Conference on Research \& Development in Information
  Retrieval}}. \bibinfo{pages}{1253--1256}.
\newblock


\bibitem[\protect\citeauthoryear{Litschko, Glava\v{s}, Vulic, and
  Dietz}{Litschko et~al\mbox{.}}{2019}]%
        {10.1145/3331184.3331324}
\bibfield{author}{\bibinfo{person}{Robert Litschko}, \bibinfo{person}{Goran
  Glava\v{s}}, \bibinfo{person}{Ivan Vulic}, {and} \bibinfo{person}{Laura
  Dietz}.} \bibinfo{year}{2019}\natexlab{}.
\newblock \showarticletitle{Evaluating Resource-Lean Cross-Lingual Embedding
  Models in Unsupervised Retrieval}. In \bibinfo{booktitle}{\emph{Proceedings
  of the 42nd International ACM SIGIR Conference on Research and Development in
  Information Retrieval}} (Paris, France) \emph{(\bibinfo{series}{SIGIR'19})}.
  \bibinfo{publisher}{Association for Computing Machinery},
  \bibinfo{address}{New York, NY, USA}, \bibinfo{pages}{1109–1112}.
\newblock
\showISBNx{9781450361729}
\urldef\tempurl%
\url{https://doi.org/10.1145/3331184.3331324}
\showDOI{\tempurl}


\bibitem[\protect\citeauthoryear{MacAvaney, Yates, Cohan, and
  Goharian}{MacAvaney et~al\mbox{.}}{2019}]%
        {macavaney2019cedr}
\bibfield{author}{\bibinfo{person}{Sean MacAvaney}, \bibinfo{person}{Andrew
  Yates}, \bibinfo{person}{Arman Cohan}, {and} \bibinfo{person}{Nazli
  Goharian}.} \bibinfo{year}{2019}\natexlab{}.
\newblock \showarticletitle{CEDR: Contextualized embeddings for document
  ranking}. In \bibinfo{booktitle}{\emph{Proceedings of the 42nd International
  ACM SIGIR Conference on Research and Development in Information Retrieval}}.
  \bibinfo{pages}{1101--1104}.
\newblock


\bibitem[\protect\citeauthoryear{Nogueira and Cho}{Nogueira and Cho}{2019}]%
        {nogueira2019passage}
\bibfield{author}{\bibinfo{person}{Rodrigo Nogueira} {and}
  \bibinfo{person}{Kyunghyun Cho}.} \bibinfo{year}{2019}\natexlab{}.
\newblock \showarticletitle{Passage Re-ranking with BERT}.
\newblock \bibinfo{journal}{\emph{arXiv preprint arXiv:1901.04085}}
  (\bibinfo{year}{2019}).
\newblock


\bibitem[\protect\citeauthoryear{Och and Ney}{Och and Ney}{2003}]%
        {och2003systematic}
\bibfield{author}{\bibinfo{person}{Franz~Josef Och} {and}
  \bibinfo{person}{Hermann Ney}.} \bibinfo{year}{2003}\natexlab{}.
\newblock \showarticletitle{A systematic comparison of various statistical
  alignment models}.
\newblock \bibinfo{journal}{\emph{Computational linguistics}}
  \bibinfo{volume}{29}, \bibinfo{number}{1} (\bibinfo{year}{2003}),
  \bibinfo{pages}{19--51}.
\newblock


\bibitem[\protect\citeauthoryear{Peters}{Peters}{2005}]%
        {10.1007/11519645_1}
\bibfield{author}{\bibinfo{person}{Carol Peters}.}
  \bibinfo{year}{2005}\natexlab{}.
\newblock \showarticletitle{What Happened in CLEF 2004?}. In
  \bibinfo{booktitle}{\emph{Multilingual Information Access for Text, Speech
  and Images}}, \bibfield{editor}{\bibinfo{person}{Carol Peters},
  \bibinfo{person}{Paul Clough}, \bibinfo{person}{Julio Gonzalo},
  \bibinfo{person}{Gareth J.~F. Jones}, \bibinfo{person}{Michael Kluck}, {and}
  \bibinfo{person}{Bernardo Magnini}} (Eds.). \bibinfo{publisher}{Springer
  Berlin Heidelberg}, \bibinfo{address}{Berlin, Heidelberg},
  \bibinfo{pages}{1--9}.
\newblock
\showISBNx{978-3-540-32051-7}


\bibitem[\protect\citeauthoryear{Peters}{Peters}{2006}]%
        {10.1007/11878773_1}
\bibfield{author}{\bibinfo{person}{Carol Peters}.}
  \bibinfo{year}{2006}\natexlab{}.
\newblock \showarticletitle{What Happened in CLEF 2005}. In
  \bibinfo{booktitle}{\emph{Accessing Multilingual Information Repositories}},
  \bibfield{editor}{\bibinfo{person}{Carol Peters}, \bibinfo{person}{Fredric~C.
  Gey}, \bibinfo{person}{Julio Gonzalo}, \bibinfo{person}{Henning M{\"u}ller},
  \bibinfo{person}{Gareth J.~F. Jones}, \bibinfo{person}{Michael Kluck},
  \bibinfo{person}{Bernardo Magnini}, {and} \bibinfo{person}{Maarten de~Rijke}}
  (Eds.). \bibinfo{publisher}{Springer Berlin Heidelberg},
  \bibinfo{address}{Berlin, Heidelberg}, \bibinfo{pages}{1--10}.
\newblock
\showISBNx{978-3-540-45700-8}


\bibitem[\protect\citeauthoryear{Peters}{Peters}{2007}]%
        {10.1007/978-3-540-74999-8_1}
\bibfield{author}{\bibinfo{person}{Carol Peters}.}
  \bibinfo{year}{2007}\natexlab{}.
\newblock \showarticletitle{What Happened in CLEF 2006}. In
  \bibinfo{booktitle}{\emph{Evaluation of Multilingual and Multi-modal
  Information Retrieval}}, \bibfield{editor}{\bibinfo{person}{Carol Peters},
  \bibinfo{person}{Paul Clough}, \bibinfo{person}{Fredric~C. Gey},
  \bibinfo{person}{Jussi Karlgren}, \bibinfo{person}{Bernardo Magnini},
  \bibinfo{person}{Douglas~W. Oard}, \bibinfo{person}{Maarten de~Rijke}, {and}
  \bibinfo{person}{Maximilian Stempfhuber}} (Eds.).
  \bibinfo{publisher}{Springer Berlin Heidelberg}, \bibinfo{address}{Berlin,
  Heidelberg}, \bibinfo{pages}{1--10}.
\newblock
\showISBNx{978-3-540-74999-8}


\bibitem[\protect\citeauthoryear{Peters}{Peters}{2008}]%
        {10.1007/978-3-540-85760-0_1}
\bibfield{author}{\bibinfo{person}{Carol Peters}.}
  \bibinfo{year}{2008}\natexlab{}.
\newblock \showarticletitle{What Happened in CLEF 2007}. In
  \bibinfo{booktitle}{\emph{Advances in Multilingual and Multimodal Information
  Retrieval}}, \bibfield{editor}{\bibinfo{person}{Carol Peters},
  \bibinfo{person}{Valentin Jijkoun}, \bibinfo{person}{Thomas Mandl},
  \bibinfo{person}{Henning M{\"u}ller}, \bibinfo{person}{Douglas~W. Oard},
  \bibinfo{person}{Anselmo Pe{\~{n}}as}, \bibinfo{person}{Vivien Petras}, {and}
  \bibinfo{person}{Diana Santos}} (Eds.). \bibinfo{publisher}{Springer Berlin
  Heidelberg}, \bibinfo{address}{Berlin, Heidelberg}, \bibinfo{pages}{1--12}.
\newblock
\showISBNx{978-3-540-85760-0}


\bibitem[\protect\citeauthoryear{Peters}{Peters}{2009}]%
        {10.1007/978-3-642-04447-2_1}
\bibfield{author}{\bibinfo{person}{Carol Peters}.}
  \bibinfo{year}{2009}\natexlab{}.
\newblock \showarticletitle{What Happened in CLEF 2008}. In
  \bibinfo{booktitle}{\emph{Evaluating Systems for Multilingual and Multimodal
  Information Access}}, \bibfield{editor}{\bibinfo{person}{Carol Peters},
  \bibinfo{person}{Thomas Deselaers}, \bibinfo{person}{Nicola Ferro},
  \bibinfo{person}{Julio Gonzalo}, \bibinfo{person}{Gareth J.~F. Jones},
  \bibinfo{person}{Mikko Kurimo}, \bibinfo{person}{Thomas Mandl},
  \bibinfo{person}{Anselmo Pe{\~{n}}as}, {and} \bibinfo{person}{Vivien Petras}}
  (Eds.). \bibinfo{publisher}{Springer Berlin Heidelberg},
  \bibinfo{address}{Berlin, Heidelberg}, \bibinfo{pages}{1--14}.
\newblock
\showISBNx{978-3-642-04447-2}


\bibitem[\protect\citeauthoryear{Peters, Neumann, RobertLLogan, Schwartz,
  Joshi, Singh, and Smith}{Peters et~al\mbox{.}}{2019}]%
        {Peters2019KnowledgeEC}
\bibfield{author}{\bibinfo{person}{Matthew~E. Peters}, \bibinfo{person}{Mark
  Neumann}, \bibinfo{person}{IV RobertLLogan}, \bibinfo{person}{Roy Schwartz},
  \bibinfo{person}{V. Joshi}, \bibinfo{person}{Sameer Singh}, {and}
  \bibinfo{person}{Noah~A. Smith}.} \bibinfo{year}{2019}\natexlab{}.
\newblock \showarticletitle{Knowledge Enhanced Contextual Word
  Representations}. In \bibinfo{booktitle}{\emph{EMNLP/IJCNLP}}.
\newblock


\bibitem[\protect\citeauthoryear{Pires, Schlinger, and Garrette}{Pires
  et~al\mbox{.}}{2019}]%
        {pires-etal-2019-multilingual}
\bibfield{author}{\bibinfo{person}{Telmo Pires}, \bibinfo{person}{Eva
  Schlinger}, {and} \bibinfo{person}{Dan Garrette}.}
  \bibinfo{year}{2019}\natexlab{}.
\newblock \showarticletitle{How Multilingual is Multilingual {BERT}?}. In
  \bibinfo{booktitle}{\emph{Proceedings of the 57th Annual Meeting of the
  Association for Computational Linguistics}}. \bibinfo{publisher}{Association
  for Computational Linguistics}, \bibinfo{address}{Florence, Italy},
  \bibinfo{pages}{4996--5001}.
\newblock
\urldef\tempurl%
\url{https://doi.org/10.18653/v1/P19-1493}
\showDOI{\tempurl}


\bibitem[\protect\citeauthoryear{Qiao, Xiong, Liu, and Liu}{Qiao
  et~al\mbox{.}}{2019}]%
        {qiao2019understanding}
\bibfield{author}{\bibinfo{person}{Yifan Qiao}, \bibinfo{person}{Chenyan
  Xiong}, \bibinfo{person}{Zhenghao Liu}, {and} \bibinfo{person}{Zhiyuan Liu}.}
  \bibinfo{year}{2019}\natexlab{}.
\newblock \bibinfo{title}{Understanding the Behaviors of BERT in Ranking}.
\newblock
\newblock
\showeprint[arxiv]{1904.07531}~[cs.IR]


\bibitem[\protect\citeauthoryear{Robertson, Walker, Jones, Hancock-Beaulieu,
  Gatford, et~al\mbox{.}}{Robertson et~al\mbox{.}}{1995}]%
        {robertson1995okapi}
\bibfield{author}{\bibinfo{person}{Stephen~E Robertson}, \bibinfo{person}{Steve
  Walker}, \bibinfo{person}{Susan Jones}, \bibinfo{person}{Micheline~M
  Hancock-Beaulieu}, \bibinfo{person}{Mike Gatford}, {et~al\mbox{.}}}
  \bibinfo{year}{1995}\natexlab{}.
\newblock \showarticletitle{Okapi at TREC-3}.
\newblock \bibinfo{journal}{\emph{Nist Special Publication Sp}}
  \bibinfo{volume}{109} (\bibinfo{year}{1995}), \bibinfo{pages}{109}.
\newblock


\bibitem[\protect\citeauthoryear{Saleh and Pecina}{Saleh and Pecina}{2020}]%
        {saleh-pecina-2020-document}
\bibfield{author}{\bibinfo{person}{Shadi Saleh} {and} \bibinfo{person}{Pavel
  Pecina}.} \bibinfo{year}{2020}\natexlab{}.
\newblock \showarticletitle{Document Translation vs. Query Translation for
  Cross-Lingual Information Retrieval in the Medical Domain}. In
  \bibinfo{booktitle}{\emph{Proceedings of the 58th Annual Meeting of the
  Association for Computational Linguistics}}. \bibinfo{publisher}{Association
  for Computational Linguistics}, \bibinfo{address}{Online},
  \bibinfo{pages}{6849--6860}.
\newblock
\urldef\tempurl%
\url{https://doi.org/10.18653/v1/2020.acl-main.613}
\showDOI{\tempurl}


\bibitem[\protect\citeauthoryear{Sarwar, Bonab, and Allan}{Sarwar
  et~al\mbox{.}}{2019}]%
        {sarwar-etal-2019-multi}
\bibfield{author}{\bibinfo{person}{Sheikh~Muhammad Sarwar},
  \bibinfo{person}{Hamed Bonab}, {and} \bibinfo{person}{James Allan}.}
  \bibinfo{year}{2019}\natexlab{}.
\newblock \showarticletitle{A Multi-Task Architecture on Relevance-based Neural
  Query Translation}. In \bibinfo{booktitle}{\emph{Proceedings of the 57th
  Annual Meeting of the Association for Computational Linguistics}}.
  \bibinfo{publisher}{Association for Computational Linguistics},
  \bibinfo{address}{Florence, Italy}, \bibinfo{pages}{6339--6344}.
\newblock
\urldef\tempurl%
\url{https://doi.org/10.18653/v1/P19-1639}
\showDOI{\tempurl}


\bibitem[\protect\citeauthoryear{Sasaki, Sun, Schamoni, Duh, and Inui}{Sasaki
  et~al\mbox{.}}{2018}]%
        {sasaki2018cross}
\bibfield{author}{\bibinfo{person}{Shota Sasaki}, \bibinfo{person}{Shuo Sun},
  \bibinfo{person}{Shigehiko Schamoni}, \bibinfo{person}{Kevin Duh}, {and}
  \bibinfo{person}{Kentaro Inui}.} \bibinfo{year}{2018}\natexlab{}.
\newblock \showarticletitle{Cross-lingual learning-to-rank with shared
  representations}. In \bibinfo{booktitle}{\emph{Proceedings of the 2018
  Conference of the North American Chapter of the Association for Computational
  Linguistics: Human Language Technologies, Volume 2 (Short Papers)}}.
  \bibinfo{pages}{458--463}.
\newblock


\bibitem[\protect\citeauthoryear{Schamoni, Hieber, Sokolov, and
  Riezler}{Schamoni et~al\mbox{.}}{2014}]%
        {schamoni2014learning}
\bibfield{author}{\bibinfo{person}{Shigehiko Schamoni}, \bibinfo{person}{Felix
  Hieber}, \bibinfo{person}{Artem Sokolov}, {and} \bibinfo{person}{Stefan
  Riezler}.} \bibinfo{year}{2014}\natexlab{}.
\newblock \showarticletitle{Learning translational and knowledge-based
  similarities from relevance rankings for cross-language retrieval}. In
  \bibinfo{booktitle}{\emph{Proceedings of the 52nd Annual Meeting of the
  Association for Computational Linguistics (Volume 2: Short Papers)}}.
  \bibinfo{pages}{488--494}.
\newblock


\bibitem[\protect\citeauthoryear{Tenney, Das, and Pavlick}{Tenney
  et~al\mbox{.}}{2019}]%
        {tenney-etal-2019-bert}
\bibfield{author}{\bibinfo{person}{Ian Tenney}, \bibinfo{person}{Dipanjan Das},
  {and} \bibinfo{person}{Ellie Pavlick}.} \bibinfo{year}{2019}\natexlab{}.
\newblock \showarticletitle{{BERT} Rediscovers the Classical {NLP} Pipeline}.
  In \bibinfo{booktitle}{\emph{Proceedings of the 57th Annual Meeting of the
  Association for Computational Linguistics}}. \bibinfo{publisher}{Association
  for Computational Linguistics}, \bibinfo{address}{Florence, Italy},
  \bibinfo{pages}{4593--4601}.
\newblock
\urldef\tempurl%
\url{https://doi.org/10.18653/v1/P19-1452}
\showDOI{\tempurl}


\bibitem[\protect\citeauthoryear{Vaswani, Shazeer, Parmar, Uszkoreit, Jones,
  Gomez, Kaiser, and Polosukhin}{Vaswani et~al\mbox{.}}{2017}]%
        {vaswani2017attention}
\bibfield{author}{\bibinfo{person}{Ashish Vaswani}, \bibinfo{person}{Noam
  Shazeer}, \bibinfo{person}{Niki Parmar}, \bibinfo{person}{Jakob Uszkoreit},
  \bibinfo{person}{Llion Jones}, \bibinfo{person}{Aidan~N Gomez},
  \bibinfo{person}{Lukasz Kaiser}, {and} \bibinfo{person}{Illia Polosukhin}.}
  \bibinfo{year}{2017}\natexlab{}.
\newblock \showarticletitle{Attention is all you need}.
\newblock \bibinfo{journal}{\emph{arXiv preprint arXiv:1706.03762}}
  (\bibinfo{year}{2017}).
\newblock


\bibitem[\protect\citeauthoryear{Vuli{\'c} and Moens}{Vuli{\'c} and
  Moens}{2015}]%
        {vulic2015monolingual}
\bibfield{author}{\bibinfo{person}{Ivan Vuli{\'c}} {and}
  \bibinfo{person}{Marie-Francine Moens}.} \bibinfo{year}{2015}\natexlab{}.
\newblock \showarticletitle{Monolingual and cross-lingual information retrieval
  models based on (bilingual) word embeddings}. In
  \bibinfo{booktitle}{\emph{Proceedings of the 38th international ACM SIGIR
  conference on research and development in information retrieval}}.
  \bibinfo{pages}{363--372}.
\newblock


\bibitem[\protect\citeauthoryear{Xia, Wang, Tian, and Chang}{Xia
  et~al\mbox{.}}{2021}]%
        {attention_sts}
\bibfield{author}{\bibinfo{person}{Tingyu Xia}, \bibinfo{person}{Yue Wang},
  \bibinfo{person}{Yuan Tian}, {and} \bibinfo{person}{Yi Chang}.}
  \bibinfo{year}{2021}\natexlab{}.
\newblock \showarticletitle{Using Prior Knowledge to Guide BERT's Attention in
  Semantic Textual Matching Tasks}.
\newblock  (\bibinfo{year}{2021}).
\newblock


\bibitem[\protect\citeauthoryear{Xiong, Du, Wang, and Stoyanov}{Xiong
  et~al\mbox{.}}{2020}]%
        {Xiong2020PretrainedEW}
\bibfield{author}{\bibinfo{person}{Wenhan Xiong}, \bibinfo{person}{Jingfei Du},
  \bibinfo{person}{William~Yang Wang}, {and} \bibinfo{person}{Veselin
  Stoyanov}.} \bibinfo{year}{2020}\natexlab{}.
\newblock \showarticletitle{Pretrained Encyclopedia: Weakly Supervised
  Knowledge-Pretrained Language Model}.
\newblock \bibinfo{journal}{\emph{ArXiv}}  \bibinfo{volume}{abs/1912.09637}
  (\bibinfo{year}{2020}).
\newblock


\bibitem[\protect\citeauthoryear{Yarmohammadi, Ma, Hisamoto, Rahman, Wang, Xu,
  Povey, Koehn, and Duh}{Yarmohammadi et~al\mbox{.}}{2019}]%
        {yarmohammadi-etal-2019-robust}
\bibfield{author}{\bibinfo{person}{Mahsa Yarmohammadi}, \bibinfo{person}{Xutai
  Ma}, \bibinfo{person}{Sorami Hisamoto}, \bibinfo{person}{Muhammad Rahman},
  \bibinfo{person}{Yiming Wang}, \bibinfo{person}{Hainan Xu},
  \bibinfo{person}{Daniel Povey}, \bibinfo{person}{Philipp Koehn}, {and}
  \bibinfo{person}{Kevin Duh}.} \bibinfo{year}{2019}\natexlab{}.
\newblock \showarticletitle{Robust Document Representations for Cross-Lingual
  Information Retrieval in Low-Resource Settings}. In
  \bibinfo{booktitle}{\emph{Proceedings of Machine Translation Summit XVII
  Volume 1: Research Track}}. \bibinfo{publisher}{European Association for
  Machine Translation}, \bibinfo{address}{Dublin, Ireland},
  \bibinfo{pages}{12--20}.
\newblock
\urldef\tempurl%
\url{https://www.aclweb.org/anthology/W19-6602}
\showURL{%
\tempurl}


\bibitem[\protect\citeauthoryear{Yates, Nogueira, and Lin}{Yates
  et~al\mbox{.}}{2021}]%
        {yates-etal-2021-pretrained}
\bibfield{author}{\bibinfo{person}{Andrew Yates}, \bibinfo{person}{Rodrigo
  Nogueira}, {and} \bibinfo{person}{Jimmy Lin}.}
  \bibinfo{year}{2021}\natexlab{}.
\newblock \showarticletitle{Pretrained Transformers for Text Ranking: {BERT}
  and Beyond}. In \bibinfo{booktitle}{\emph{Proceedings of the 2021 Conference
  of the North American Chapter of the Association for Computational
  Linguistics: Human Language Technologies: Tutorials}}.
  \bibinfo{publisher}{Association for Computational Linguistics},
  \bibinfo{address}{Online}, \bibinfo{pages}{1--4}.
\newblock
\urldef\tempurl%
\url{https://www.aclweb.org/anthology/2021.naacl-tutorials.1}
\showURL{%
\tempurl}


\bibitem[\protect\citeauthoryear{Yu and Allan}{Yu and Allan}{2020}]%
        {10.1145/3397271.3401322}
\bibfield{author}{\bibinfo{person}{Puxuan Yu} {and} \bibinfo{person}{James
  Allan}.} \bibinfo{year}{2020}\natexlab{}.
\newblock \bibinfo{booktitle}{\emph{A Study of Neural Matching Models for
  Cross-Lingual IR}}.
\newblock \bibinfo{publisher}{Association for Computing Machinery},
  \bibinfo{address}{New York, NY, USA}, \bibinfo{pages}{1637–1640}.
\newblock
\showISBNx{9781450380164}
\urldef\tempurl%
\url{https://doi.org/10.1145/3397271.3401322}
\showURL{%
\tempurl}


\bibitem[\protect\citeauthoryear{Yu, Fei, and Li}{Yu et~al\mbox{.}}{2021}]%
        {10.1145/3442381.3449830}
\bibfield{author}{\bibinfo{person}{Puxuan Yu}, \bibinfo{person}{Hongliang Fei},
  {and} \bibinfo{person}{Ping Li}.} \bibinfo{year}{2021}\natexlab{}.
\newblock \showarticletitle{Cross-Lingual Language Model Pretraining for
  Retrieval}. In \bibinfo{booktitle}{\emph{Proceedings of the Web Conference
  2021}} (Ljubljana, Slovenia) \emph{(\bibinfo{series}{WWW '21})}.
  \bibinfo{publisher}{Association for Computing Machinery},
  \bibinfo{address}{New York, NY, USA}, \bibinfo{pages}{1029–1039}.
\newblock
\showISBNx{9781450383127}
\urldef\tempurl%
\url{https://doi.org/10.1145/3442381.3449830}
\showDOI{\tempurl}


\bibitem[\protect\citeauthoryear{Zhan, Mao, Liu, Zhang, and Ma}{Zhan
  et~al\mbox{.}}{2020}]%
        {10.1145/3397271.3401325}
\bibfield{author}{\bibinfo{person}{Jingtao Zhan}, \bibinfo{person}{Jiaxin Mao},
  \bibinfo{person}{Yiqun Liu}, \bibinfo{person}{Min Zhang}, {and}
  \bibinfo{person}{Shaoping Ma}.} \bibinfo{year}{2020}\natexlab{}.
\newblock \showarticletitle{An Analysis of BERT in Document Ranking}. In
  \bibinfo{booktitle}{\emph{Proceedings of the 43rd International ACM SIGIR
  Conference on Research and Development in Information Retrieval}} (Virtual
  Event, China) \emph{(\bibinfo{series}{SIGIR '20})}.
  \bibinfo{publisher}{Association for Computing Machinery},
  \bibinfo{address}{New York, NY, USA}, \bibinfo{pages}{1941–1944}.
\newblock
\showISBNx{9781450380164}
\urldef\tempurl%
\url{https://doi.org/10.1145/3397271.3401325}
\showDOI{\tempurl}


\bibitem[\protect\citeauthoryear{Zhang, Karakos, Hartmann, Srivastava, Tarlin,
  Akodes, Gouda, Bathool, Zhao, Jiang, Schwartz, and Makhoul}{Zhang
  et~al\mbox{.}}{2020}]%
        {zhang-etal-2020-2019}
\bibfield{author}{\bibinfo{person}{Le Zhang}, \bibinfo{person}{Damianos
  Karakos}, \bibinfo{person}{William Hartmann}, \bibinfo{person}{Manaj
  Srivastava}, \bibinfo{person}{Lee Tarlin}, \bibinfo{person}{David Akodes},
  \bibinfo{person}{Sanjay~Krishna Gouda}, \bibinfo{person}{Numra Bathool},
  \bibinfo{person}{Lingjun Zhao}, \bibinfo{person}{Zhuolin Jiang},
  \bibinfo{person}{Richard Schwartz}, {and} \bibinfo{person}{John Makhoul}.}
  \bibinfo{year}{2020}\natexlab{}.
\newblock \showarticletitle{The 2019 {BBN} Cross-lingual Information Retrieval
  System}. In \bibinfo{booktitle}{\emph{Proceedings of the workshop on
  Cross-Language Search and Summarization of Text and Speech (CLSSTS2020)}}.
  \bibinfo{publisher}{European Language Resources Association},
  \bibinfo{address}{Marseille, France}, \bibinfo{pages}{44--51}.
\newblock
\showISBNx{979-10-95546-55-9}
\urldef\tempurl%
\url{https://www.aclweb.org/anthology/2020.clssts-1.8}
\showURL{%
\tempurl}


\bibitem[\protect\citeauthoryear{Zhang, Han, Liu, Jiang, Sun, and Liu}{Zhang
  et~al\mbox{.}}{2019}]%
        {zhang-etal-2019-ernie}
\bibfield{author}{\bibinfo{person}{Zhengyan Zhang}, \bibinfo{person}{Xu Han},
  \bibinfo{person}{Zhiyuan Liu}, \bibinfo{person}{Xin Jiang},
  \bibinfo{person}{Maosong Sun}, {and} \bibinfo{person}{Qun Liu}.}
  \bibinfo{year}{2019}\natexlab{}.
\newblock \showarticletitle{{ERNIE}: Enhanced Language Representation with
  Informative Entities}. In \bibinfo{booktitle}{\emph{Proceedings of the 57th
  Annual Meeting of the Association for Computational Linguistics}}.
  \bibinfo{publisher}{Association for Computational Linguistics},
  \bibinfo{address}{Florence, Italy}, \bibinfo{pages}{1441--1451}.
\newblock
\urldef\tempurl%
\url{https://doi.org/10.18653/v1/P19-1139}
\showDOI{\tempurl}


\bibitem[\protect\citeauthoryear{Zhao, Zbib, Jiang, Karakos, and Huang}{Zhao
  et~al\mbox{.}}{2019}]%
        {zhao2019weakly}
\bibfield{author}{\bibinfo{person}{Lingjun Zhao}, \bibinfo{person}{Rabih Zbib},
  \bibinfo{person}{Zhuolin Jiang}, \bibinfo{person}{Damianos Karakos}, {and}
  \bibinfo{person}{Zhongqiang Huang}.} \bibinfo{year}{2019}\natexlab{}.
\newblock \showarticletitle{Weakly supervised attentional model for low
  resource ad-hoc cross-lingual information retrieval}. In
  \bibinfo{booktitle}{\emph{Proceedings of the 2nd Workshop on Deep Learning
  Approaches for Low-Resource NLP (DeepLo 2019)}}. \bibinfo{pages}{259--264}.
\newblock


\bibitem[\protect\citeauthoryear{Zhao and Bethard}{Zhao and Bethard}{2020}]%
        {zhao-bethard-2020-berts}
\bibfield{author}{\bibinfo{person}{Yiyun Zhao} {and} \bibinfo{person}{Steven
  Bethard}.} \bibinfo{year}{2020}\natexlab{}.
\newblock \showarticletitle{How does {BERT}{'}s attention change when you
  fine-tune? An analysis methodology and a case study in negation scope}. In
  \bibinfo{booktitle}{\emph{Proceedings of the 58th Annual Meeting of the
  Association for Computational Linguistics}}. \bibinfo{publisher}{Association
  for Computational Linguistics}, \bibinfo{address}{Online},
  \bibinfo{pages}{4729--4747}.
\newblock
\urldef\tempurl%
\url{https://doi.org/10.18653/v1/2020.acl-main.429}
\showDOI{\tempurl}


\end{thebibliography}

\end{document}